\documentclass[runningheads]{llncs}

\usepackage[table]{xcolor}
\usepackage{orcidlink}
\usepackage{listings}
\usepackage{caption}
\usepackage{courier}
\usepackage{tabularx}
\usepackage{multirow}
\usepackage{booktabs}
\usepackage{makecell}
\usepackage{array}
\usepackage{ragged2e}
\usepackage{makecell}
\usepackage{cite}
\usepackage{amsmath,amssymb,amsfonts}
\usepackage{pifont}
\newcommand{\cmark}{\ding{51}}  
\newcommand{\xmark}{\ding{55}}  
\usepackage{algorithmic}
\usepackage{graphicx}
\usepackage{textcomp}
\makeatletter
\def\input@path{{./}}
\makeatother
\usepackage[most]{tcolorbox}
\usetikzlibrary{calc}
\usepackage{tikz}
\usetikzlibrary{arrows.meta, positioning}
\usepackage{listings}

\lstdefinestyle{jsonstyle}{
	basicstyle=\ttfamily\small,
	breaklines=true,
	breakatwhitespace=true,
	showstringspaces=false,
	frame=none,
	columns=fullflexible,
	keepspaces=true,
	escapeinside={\%\%}{\%\%},
	stringstyle=\color{brown},
	keywordstyle=\color{blue},
	commentstyle=\color{gray}
}

\def\BibTeX{{\rm B\kern-.05em{\sc i\kern-.025em b}\kern-.08em
    T\kern-.1667em\lower.7ex\hbox{E}\kern-.125emX}}

\begin{document}

\title{PenTest2.0: Towards Autonomous Privilege Escalation Using GenAI}

\titlerunning{Introducing PenTest2.0}

\author{
 Haitham S.\ Al-Sinani\inst{1}\orcidlink{0009-0005-0453-3335}
\and Chris J.\ Mitchell\inst{2}\orcidlink{0000-0002-6118-0055}
}

\institute{
Department of Cybersecurity and Quality Assurance, Diwan of Royal Court,  Muscat, Oman,
\email{hsssinani@diwan.gov.om} \\
\and Department of Information Security, Royal Holloway, University of London, Egham, UK.
\email{C.Mitchell@rhul.ac.uk}\\
 }
\authorrunning{H. Al-Sinani \& C. Mitchell}

\maketitle

\begin{abstract}
Ethical hacking today relies on highly skilled practitioners executing complex sequences of
commands, which is inherently time-consuming, difficult to scale, and prone to human error. To help
mitigate these limitations, we previously introduced \texttt{PenTest++}, an AI-augmented system
combining automation with generative AI supporting ethical hacking workflows. However, a key
limitation of \texttt{PenTest++} was its lack of support for \texttt{privilege escalation}, a crucial
element of ethical hacking. In this paper we present \texttt{PenTest2.0}, a substantial evolution
of \texttt{PenTest++} supporting automated privilege escalation driven entirely by Large Language
Model reasoning. It also incorporates several significant enhancements: \texttt{Retrieval-Augmented
Generation}, including both one-line and offline modes; \texttt{Chain-of-Thought}  prompting for
intermediate reasoning; persistent \texttt{PenTest Task Trees} to track goal progression across
turns; and the optional integration of human-authored hints. We describe how it operates, present a
proof-of-concept prototype, and discuss its benefits and limitations. We also describe application
of the system to a controlled Linux target, showing it can carry out multi-turn, adaptive privilege
escalation. We explain the rationale behind its core design choices, and provide comprehensive
testing results and cost analysis. Our findings indicate that \texttt{PenTest2.0} represents a
meaningful step toward practical, scalable, AI-automated penetration testing, whilst highlighting
the shortcomings of generative AI systems, particularly their sensitivity to prompt structure,
execution context, and semantic drift — reinforcing the need for further research and refinement in this emerging space.

\end{abstract}

\section{Introduction}
\label{Introduction}

Penetration testing~\cite{NIST800-115} (PenTesting) is a cornerstone of modern cybersecurity practice, enabling organisations to identify and remediate vulnerabilities before they are exploited by malicious actors. Yet, despite its critical importance, the process of ethical hacking remains heavily reliant on individual expertise. Professionals must craft and execute complex chains of commands, interpret system feedback, and iteratively adapt their tactics — all under time constraints and with potential for error. These challenges make PenTesting inherently labour-intensive, costly, and difficult to scale.

To address some of these limitations, we previously introduced \texttt{PenTest++}~\cite{2025_PenTest++_HC_CyBAI}, an AI-augmented tool that leverages GenAI to assist in automating core PenTesting activities such as reconnaissance, scanning,
enumeration, exploitation, and documentation, while preserving user control and adaptability. By
embedding generative AI (GenAI) into the ethical hacking workflow, \texttt{PenTest++} successfully reduced
cognitive load and enabled context-aware suggestions. However, it lacked support for one of the
most technically demanding and critical phases in the PenTesting lifecycle: \texttt{privilege escalation}  PrivEsc. This
phase involves exploiting misconfigurations or software vulnerabilities to elevate a compromised
user account’s privileges — typically from a low-privileged  user to administrative or
root-level access. PrivEsc is essential for expanding access, simulating realistic attack
scenarios, and achieving full control over the target system. Without it, the scope and
effectiveness of any automated PenTesting system remain inherently constrained.

Furthermore, \texttt{PenTest++} did not incorporate  reasoning and tracking capabilities such as Retrieval-Augmented Generation (RAG), Chain-of-Thought (CoT) prompting, PenTest Task Trees (PTTs), or user-injected hints. To address these limitations, and in addition to enabling automated PrivEsc, we present \texttt{PenTest2.0} — a significantly enhanced version of our earlier system. \texttt{PenTest2.0} introduces several architectural and functional improvements aimed at deepening reasoning, improving traceability, and increasing autonomy. These include support for RAG in both online and offline modes; CoT prompting to facilitate intermediate reasoning steps; PTTs as a persistent task structure for managing goals and subtasks across reasoning turns\footnote{For the purposes of this paper a ‘turn’ is one automatically-executed
	prompt–response exchange between PenTest2.0 and the LLM.}; and the integration of optional human-authored hints to guide the Large Language Model (LLM) behaviour where necessary. The system also logs executed commands, token usage, cost estimates, reasoning outputs, and failure diagnostics to enable detailed post-run analysis and system introspection.

We evaluated \texttt{PenTest2.0} on a controlled Linux testbed using an assumed-breach scenario, a common and well-established setup in PenTesting research~\cite{2023_GettingGettingPwndbyAI_PenTestingWithLLMs_Happe}. 
Our results show that \texttt{PenTest2.0} marks a meaningful advancement toward scalable, realistic, and agentic AI-driven PenTesting. At the same time, the evaluation also surfaces key limitations — such as sensitivity to prompt formulation and command execution fragility — which highlight the need for continued research and refinement in this emerging domain.

The remainder of this paper is structured as follows.
Section~\ref{Research Questions and Contributions} defines  the research questions and outlines the key contributions of this work.
Section~\ref{Background} reviews recent advances, including CoT, RAG, and PTT, which inform our approach.
Section~\ref{PenTest2.0 Operation} describes how \texttt{PenTest2.0} operates, while Section~\ref{Design Choices: Discussion and Rationale} explains the rationale behind the core system design choices.
Section~\ref{Prototype Implementation} provides details of the prototype implementation.
Sections~\ref{Testing Results} and   \ref{sec:cost-analysis} report the testing results and cost analysis, respectively.
Section~\ref{GeneralDiscussion} discusses the broader implications, potential risks, and limitations of the study.
Section~\ref{Related work} surveys related literature.
Finally, Section~\ref{ConclusionsAndFurtherResearch} concludes the paper and outlines directions for future research.

 \section{Research Questions and Contributions}
  \label{Research Questions and Contributions}

In this work, we build upon our previously proposed system, \texttt{PenTest++}, by addressing one of its most critical limitations: the lack of support for automated PrivEsc. In extending the system to handle this key PenTesting phase, several new technical and conceptual challenges arise — particularly around multi-turn LLM reasoning, execution reliability, and system autonomy. Accordingly, our research is guided by the following central questions:

\begin{itemize}
    \item \textbf{RQ1:} To what extent can  GenAI  systems autonomously perform PrivEsc in a post-exploitation scenario under human supervision?
    \item \textbf{RQ2:} How can the integration of techniques such as RAG, CoT  prompting, PTTs, and optional human hints improve the effectiveness, reasoning depth, and traceability of GenAI-driven PenTesting?
    \item \textbf{RQ3:} What practical limitations arise when applying LLMs to live, command-executing PrivEsc tasks in real-world-like environments, and how do these limitations manifest during execution?
\end{itemize}

To explore these questions, we present the design, implementation, and evaluation of \texttt{PenTest2.0}, a significantly enhanced improvement to our earlier system. The key contributions of this paper are as follows.

\begin{itemize}
    \item \textbf{C1:} We demonstrate the \textbf{feasibility of automating} the PrivEsc phase of the ethical hacking process through an AI-augmented tool, \textbf{while putting the user in control} for safety and ethical reasons.

    \item \textbf{C2:} We introduce  a GenAI-powered,  automated \textbf{proof-of-concept prototype} for PrivEsc, enabling the system to iteratively reason, generate, execute, and adapt commands in a live post-exploitation setting — with the full intention of making it open source on GitHub.

    \item \textbf{C3:} We integrate a suite of optional, advanced features — including \textbf{RAG, CoT, PTTs,} and a lightweight \textbf{hint mechanism} — to support deeper reasoning, traceable execution, and adaptive task management across multiple LLM turns.

    \item \textbf{C4:} We conduct a comprehensive \textbf{experimental evaluation} on a realistic Linux target machine, demonstrating the tool's ability to autonomously perform PrivEsc with minimal human oversight, while systematically capturing failure modes and runtime behaviour.

    \item \textbf{C5:} We offer a \textbf{balanced and critical analysis} of \texttt{PenTest2.0}’s capabilities and limitations, identifying open challenges and proposing concrete avenues for future research aimed at advancing agentic, ethical, and dependable GenAI-assisted PenTesting.

\end{itemize}

\section{Background}
\label{Background}
This section outlines the core prompting strategies and optional enhancements underpinning \texttt{PenTest2.0}, divided into five components: the base prompt,  CoT  reasoning, Human Hint injection, RAG, and PTT-based task tracking.

\subsection{Base Prompt: Setting the Rules of Engagement}
\label{Background:BasePrompt}

The base prompt defines the operating context, rules of engagement, and input/output structure for the LLM. It establishes the simulated environment as a real-time penetration test and enforces strict behavioural boundaries — e.g., non-interactive commands only, respect for output structure, JSON compliance, and clarity of reasoning.

An \textbf{excerpt} from our base prompt includes:
	\begin{tcolorbox}[enhanced, breakable,
		colback=gray!5,               
		colframe=black!60,            
		colbacktitle=gray!20,         
		coltitle=black,               
		fonttitle=\bfseries\large,    
		title={\centering Excerpt from Our Base Prompt}, 
		boxed title style={
			colframe=black!75,
			colback=gray!20,            
			boxrule=0.5pt,
			arc=2pt,
			left=2pt, right=2pt,
			top=1pt, bottom=1pt
		},
		boxrule=0.4pt,
		arc=3pt,
		before skip=10pt, after skip=10pt,
		width=\textwidth]
	\small
	\texttt{You are assisting with Linux Privilege Escalation. You have a low-level privileged account ({{USERNAME}}), and your ultimate goal is to elevate your privilege access to root level.
	We will interact with you turn by turn: at each step, you’ll receive a summary of the target system context and you must suggest the next best command. If root is not yet achieved, we will send you the execution output. Your goal is to gain root access within a maximum of {{MAX\_TURNS}} turns.
	\\
	...
	\\
	\textbf{Rules You MUST Follow:}\\
	1. NEVER repeat a command that already succeeded *within the current run*. You MAY suggest it again in future sessions if the target system appears similar or identical. \\
	2. You MUST retry a corrected version of a previous command that failed due to syntax errors or misconfiguration. Do not suggest the exact same faulty command again. \\
	3. NEVER suggest dangerous or destructive commands, including: \\
	- rm -rf *, rm -rf /, dd if=/dev/zero \\
	- zip -r backup.zip /, zip -r backup.zip /etc \\
	- Any command involving heavy backups or full system traversal \\
	4. NEVER break the JSON structure, even if unsure. \\
	\\
	...}
\end{tcolorbox}
The base prompt also specifies a \textbf{mandatory} output format, requiring the LLM to return a single valid and compact JSON object only, as follows. 
{\scriptsize
\begin{verbatim}
	{
		"command_non_interactive": "string, safe for automated execution (no $, #, `)",
		"command_interactive": "string, interactive version if applicable, else empty",
		"system_summary": "string, max 10 very short bullet points",
		"command_history": "string (max 15 lines, summarised cleanly)",
		"rationale": "string, 1-2 sentences, explaining why this command was chosen"
		{% if rag_enabled %}
		,
		"rag_search_query": "string (max 15 words)"
		{% endif %}
		{% if ptt_enabled %}
		,
		...
		{% endif %}
	}
\end{verbatim}
}
This carefully crafted instruction ensures LLM outputs are both syntactically valid and semantically appropriate for execution, supporting reproducibility and safety.

\subsection{Chain-of-Thought Reasoning (CoT)}
\label{Background:CoT}

Chain-of-Thought (CoT) prompting is a widely studied technique for enhancing the reasoning capabilities of large language models (LLMs) by encouraging them to reason about their logic before suggesting a command~\cite{wei2022chain}. Rather than generating a final answer immediately, CoT decomposes the task into intermediate steps, fostering more deliberate and explainable behaviour. This is especially valuable in complex penetration testing scenarios where outputs must be interpreted in context, and reasoning must evolve across turns.

In \texttt{PenTest2.0}, enabling CoT appends the following directive to the LLM prompt:

\begin{quote}
 \texttt{`Think step by step. First, assess the system summary for PrivEsc paths. Then, evaluate the last command and output. Finally, decide on the most logical next command.'}
\end{quote}

This explicit instruction serves two critical purposes: (1) it prompts the model to reflect on prior outputs and command history before issuing the next command, and (2) it encourages more transparent decision-making, which can help detect semantic drift or command repetition during multi-turn reasoning.

\paragraph{Variants of CoT Prompting.} In PenTest2.0, we adopt the following CoT modes:

\begin{itemize}
	\item \textbf{Zero-shot CoT:} This is the default implementation, where only a CoT instruction is added, without examples. The LLM is simply told to `\texttt{think step by step'} with guidance contextualised to PrivEsc. This leverages the LLM’s pretraining to simulate multi-step reasoning.
	
	\item \textbf{Few-shot CoT (via training-style exemplars):} Although not fine-tuned, our prompt templates include handcrafted examples that mirror realistic CoT patterns. These training-style examples (stored in the `cot.txt` file) demonstrate how the LLM should reason, respond in JSON, and annotate rationales. They are injected only in CoT-enhanced modes and serve to reinforce structured reasoning without overloading the prompt.
	
	\item \textbf{Fine-tuned CoT:} Not used in PenTest2.0, but worth noting, fine-tuned CoT models are explicitly trained on datasets annotated with intermediate reasoning~\cite{wei2022finetuned}. Our work instead elicits similar behaviour by relying on well-engineered prompt scaffolding and examples.
\end{itemize}

\paragraph{Use of Training Examples.} As stated above, PenTest2.0 optionally embeds carefully constructed training examples into the prompt when CoT is enabled. These examples are curated from prior successful root escalation attempts and highlight how to reason about sudo rights, system misconfigurations, and failed commands. The intention is to ground the LLM’s reasoning in realistic scenarios without hardcoding command strategies.

By incorporating both zero-shot and few-shot CoT logic into the prompt structure, PenTest2.0 encourages adaptive, explainable reasoning — a critical asset in achieving root access safely and efficiently under multi-turn, HITL-guided PrivEsc loops.

\subsection{Human Hint Injection}
\label{Background:HumanHint}

In many PenTesting scenarios, the operator possesses domain expertise or prior system knowledge. The Human Hint enhancement allows structured injection of such hints into the prompt, guiding the LLM’s decision-making process. Inspired by best practices in HITL design~\cite{amershi2014power,li2025largelanguagemodelsstruggle}, this strategy combines automation with expertise.

For instance, the prompt modification may include:
\begin{quote}
	``Human Hint: Use the `id' command instead of the `/bin/sh' for root automated verification.''
\end{quote}
This balances autonomy with control, helping develop effective exploits.

\subsection{Retrieval-Augmented Generation (RAG)}
\label{Background:RAG}

 RAG  enhances the reasoning capabilities of LLMs  by grounding their responses in external knowledge sources. Rather than relying solely on internalised pretraining, RAG enables the LLM to reason in context using retrieved facts or examples relevant to the task at hand~\cite{lewis2020retrieval}. This approach helps reduce hallucinations, align model outputs with real-world constraints, and improve the quality of generated commands in complex tasks such as PrivEsc.

In \texttt{PenTest2.0}, RAG is implemented via a hybrid retrieval mechanism that supports both offline and online modes:

\begin{itemize}
	\item \textbf{Offline Mode:} During setup, the system programmatically downloads structured markdown content from trusted sources — including GTFOBins\footnote{\url{https://gtfobins.github.io/}} — and stores them in a local knowledge base. This corpus is indexed using a FAISS\footnote{FAISS (Facebook AI Similarity Search) is an open-source library for efficient similarity search and clustering of dense vectors, widely used to enable fast retrieval of semantically relevant content in AI systems and RAG setups.} vector store and queried in real time during execution. Offline RAG ensures rapid access, avoids latency, and operates independently of Internet availability. This is particularly useful in constrained or high-security environments.
	
	\item \textbf{Online Mode:} As an alternative or fallback, the system can perform live retrieval from online sources at runtime. For example, when enabled, it may query the GTFOBins website or other exploitation databases based on a search query generated by the LLM. The retrieved snippet is then parsed and injected into the next prompt, allowing the LLM to reason with up-to-date external guidance.
\end{itemize}

In both modes, retrieved content is inserted directly into the prompt in a structured format when the system  explicitly requests external help. For example:

\begin{quote}
	``Retrieved Insight: GTFOBins suggests \texttt{sudo tar} can spawn a shell using \texttt{--checkpoint-action=exec=...}.''
\end{quote}

This mechanism strengthens the LLM’s situational awareness and provides technical grounding for suggested PrivEsc commands — especially when system conditions trigger ambiguous or unexpected reasoning paths. When used effectively, RAG can bridge the gap between static prompt engineering and dynamic, context-aware AI decision-making within the PenTesting process.

\subsection{PenTest Task Tree (PTT) Tracking}
\label{Background:PTT}

Originally adopted in PenTestGPT~\cite{2024_PentestGPT_Deng}, the PTT  serves as a lightweight memory structure to preserve task structure across turns. It addresses context loss, avoids redundant suggestions, and supports tracking progress and skipped commands. 

\texttt{PenTest2.0} extends this model by allowing dynamic updates of \texttt{updated- statuses}, \texttt{new\_subtasks}, and \texttt{commands\_to\_avoid}. A sample prompt excerpt includes:
\begin{quote}
	``Current PTT Summary: Subtask 1: Examine sudo privileges. Status: pending. Subtask 2: Identify potential misconfigurations in awk.''
\end{quote}
This enables more structured exploration and improves reasoning coherence across multi-turn sessions.

\section{PenTest2.0 Operation}
\label{PenTest2.0 Operation}

\subsection{PenTest++ Operation}
\label{PenTest++ Operation}
Since \texttt{PenTest2.0} is a major extension of our previously proposed system \texttt{PenTest++}~\cite{2025_PenTest++_HC_CyBAI}, designed specifically to support GenAI-powered PrivEsc, we first summarise the key phases of \texttt{PenTest++}. This  highlights where \texttt{PenTest2.0} integrates into the broader ethical hacking workflow.
\begin{enumerate}
	\item \textbf{Reconnaissance:}
	\texttt{PenTest++} automates network discovery by executing commands to identify live systems within the target environment. 
	Users select the desired target for subsequent phases.
	
	\item \textbf{Scanning \& Enumeration:}
	\texttt{PenTest++} executes vulnerability scanning, employing tools such as \texttt{nmap} and \texttt{gobuster} to identify critical open ports, services, and misconfigurations. GenAI interprets scan outputs, correlates findings with known vulnerabilities, and provides recommendations for targeted enumeration.
	
	\item \textbf{Exploitation:}
	\texttt{PenTest++} facilitates exploitation by generating tailored payloads to exploit identified vulnerabilities, such as misconfigured services or insecure functionalities. GenAI offers strategic guidance to craft precise attack sequences, while users dynamically adjust tactics in response to real-time inputs and recommendations provided by the system.
	
	\item \textbf{Post-exploitation:}
	This is where \texttt{PenTest2.0} is activated to take over the critical task of PrivEsc. Unlike \texttt{PenTest++}, which did not support this phase, \texttt{PenTest2.0} introduces autonomous, GenAI-guided PrivEsc through an iterative command execution and feedback loop. The full operation of \texttt{PenTest2.0} is described in the following subsection.

	\item \textbf{Documentation:}
	\texttt{PenTest++} automates report generation, by leveraging GenAI  to produce a comprehensive, PenTesting  report, including logs, methodologies, key findings, and actionable recommendations. GenAI refines the structure and clarity of the documentation, ensuring it provides actionable insights for enhancing the security posture of the tested systems.
\end{enumerate}

\subsection{PenTest2.0 Operation}
\label{subPenTest2.0 Operation}
\begin{figure}[h]
	\centering
	\begin{tikzpicture}[
		node distance=1.5cm and 2cm,
		box/.style={rectangle, draw, minimum width=4.8cm, minimum height=1.1cm, align=center},
		arrow/.style={-{Latex}, thick}
		]
		
		\node[box] (builder) {\textbf{PenTest2.0} \\(Prompt Builder\\+ System State Capture\\+ Optional Enhancements)};
		\node[box, below=of builder] (promptUser) {\textbf{User} \\(Prompt Approval)};
		\node[box, below=of promptUser] (llm) {\textbf{LLM} \\(PrivEsc Command Reasoning \& Suggestion)};
		\node[box, below=of llm] (handler) {\textbf{PenTest2.0} \\(Command Handler \\+ Output Analysis\\+ Orchestrator)};
		\node[box, below=of handler] (cmdUser) {\textbf{User} \\(Command Approval)};
		\node[box, below=of cmdUser] (target) {\textbf{Target  VM}\\(Command Execution on Low-Privileged Shell)};
		\node[box, below=of target] (Outcome) {\textbf{PenTest2.0} \\(Outcome: Root or Max Turns)};
		
		\draw[arrow] (builder) -- (promptUser);
		\draw[arrow] (promptUser) -- (llm);
		\draw[arrow] (llm) -- (handler);
		\draw[arrow] (handler) -- (cmdUser);
		\draw[arrow] (cmdUser) -- (target);
		\draw[arrow] (target) -- (Outcome);
		
		\coordinate (loopDown) at ($(target.south)+(0,-1.0)$);
		\coordinate (loopRight) at ($(loopDown)+(5,0)$);
		\coordinate (loopUp) at ($(loopRight)+(0,9.2)$);
		\draw[arrow] (target.south) -- (loopDown)
		-- (loopRight)
		-- (loopUp)
		-- ($(builder.east)+(0,0)$);
		
	\end{tikzpicture}
	\caption{PenTest2.0  architecture with iterative loop}
	\label{PenTest2.0Architecture}
\end{figure}
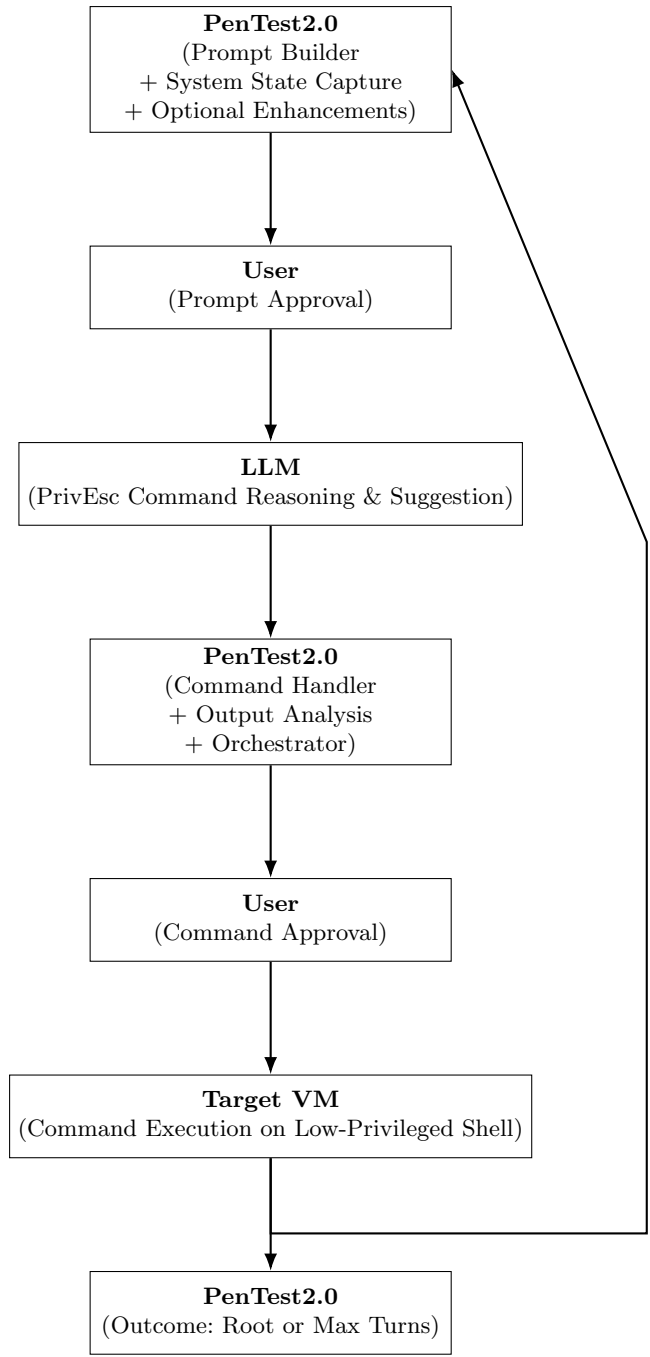

The operation of \texttt{PenTest2.0} follows a systematic methodology designed to automate PrivEsc while maintaining user control, transparency, and safety. The core loop is driven by GenAI reasoning and refined through dynamic prompt engineering, enabling multi-turn PrivEsc in realistic post-exploitation scenarios.  The operation of PenTest2.0  involves  the following key sub-phases (see Fig. \ref{PenTest2.0Architecture}).

\begin{enumerate}
	  \item \textbf{Post-Exploitation Assumption:}
	\texttt{PenTest2.0} assumes that the attacker has already obtained an initial foothold on the target machine — typically in the form of a low-privileged shell — via a preceding exploitation phase. This reflects a realistic post-exploitation scenario commonly studied in PenTesting and narrows the system’s focus to the PrivEsc phase.

	\item \textbf{Context Gathering via System Probing:}
	Before invoking the LLM, PenTest2.0 executes a predefined set of reconnaissance commands on the target to capture the system’s environment, state, and context. Common commands include \texttt{id}, \texttt{whoami}, \texttt{hostname}, \texttt{uname -a}, \texttt{sudo -l}, \texttt{env}, and \texttt{ls -la /tmp}. The outputs are aggregated, \textbf{summarised}, and used to construct the initial GenAI prompt — ensuring that the LLM begins reasoning with an accurate, cost-efficient snapshot of the target.

	\item \textbf{Initial Prompt Construction:}
	Using the captured target  state, our system dynamically constructs the first prompt by injecting relevant information into a predefined template. This prompt is then submitted to the LLM to initiate the first reasoning turn, providing the necessary context for it to suggest a suitable PrivEsc strategy.
	
	\item \textbf{Execution and Feedback Loop:}
	Upon receiving the LLM’s suggested command, \texttt{PenTest2.0} executes it on the target machine via SSH and captures the resulting output. If the command successfully yields root privileges, the process terminates. Otherwise, \texttt{PenTest2.0} constructs a new prompt, incorporating a summarised version of the previous prompt and the latest output, and sends it to the LLM for the next reasoning turn. This iterative loop continues until either root access is achieved or the maximum turn limit is reached.

	\item \textbf{User in Control:}
	Although the system is designed to enable automation, PenTest2.0 enforces user oversight at critical checkpoints. Before each prompt is sent to the LLM, the user should
	review the full prompt, token size, and estimated API cost. Similarly, each LLM-suggested command must be explicitly approved by the user before execution on the target system. This dual-approval model ensures ethical, safe operation and prevents unintended system damage or cost leakage — placing the user  in control.
	
	\item \textbf{Optional Enhancements:}
	As described in Sections \ref{Background:CoT} to \ref{Background:PTT}, PenTest2.0 supports several advanced features to enhance reasoning depth and expedite root access:
	\begin{itemize}
		\item \textbf{RAG} — in both offline and online modes — to inject relevant knowledge into LLM prompts;
		\item \textbf{CoT Prompting} to enable intermediate reasoning steps;
		\item \textbf{PTTs} to persistently track subtasks, commands, and observations across turns; and 
		\item \textbf{Hint Injection} to guide the LLM away from ineffective patterns or toward promising strategies.
	\end{itemize}
	These optional enhancements are designed to help the system reach root faster and can be toggled on or off as needed. When enabled, they are dynamically injected at runtime.
	
	 \item \textbf{Logging, Termination, and Post-Run Reporting:}
	Throughout execution, PenTest2.0 logs every prompt, command, output, token use, and root status check. Upon completion — whether due to success or max-turn exhaustion — the system generates structured reports detailing all decisions, costs, and outcomes. These logs support reproducibility, auditability, and further research.
	
\end{enumerate}

\section{Design Choices: Discussion and Rationale}
\label{Design Choices: Discussion and Rationale}

\subsection{Modular Code Architecture}

\texttt{PenTest2.0} is designed with a modular, component-based architecture that facilitates extensibility, debugging, and customisation. Core modules such as \texttt{command\_executor.py}, \texttt{llm\_connector.py}, \texttt{shell\_root\_detection.py}, and \texttt{ptt\_manager.py} encapsulate distinct responsibilities — ranging from command execution over SSH to prompt construction and root shell detection. This \texttt{separation of concerns} ensures that the system can be maintained, extended, or debugged without introducing regressions elsewhere. It also aligns with secure coding best practices by isolating sensitive operations into well-defined components.

\subsection{Dynamic Prompt Construction with Runtime Flexibility}

Unlike traditional one-shot prompting approaches, \texttt{PenTest2.0} adopts a dynamic and adaptive prompt construction strategy. Rather than relying on a monolithic prompt, the system assembles the LLM input in real time at each turn — composing it from modular building blocks that include system facts, prior outputs, task history, and optional enhancements. This composability enables more precise and cost-efficient prompt tailoring depending on the user’s goals and available resources.

To support runtime adaptability, users can toggle optional reasoning components such as  CoT, RAG, PTT, or human-injected hints, e.g. using command-line flags. When these features are disabled, \texttt{PenTest2.0} generates a minimal prompt that contains only the essential target system context, resulting in significantly lower token usage and API cost. Conversely, when deeper reasoning is required,   \texttt{PenTest2.0} supports richer prompts that embed auxiliary reasoning chains, retrieved knowledge, and strategic guidance — without needing to reconfigure or restart the system.

This runtime modularity ensures that \texttt{PenTest2.0} can operate in both budget-sensitive minimal modes and intelligence-heavy enhanced modes, offering flexibility and scalability for diverse PenTesting scenarios.

\subsection{Iterative Prompt Engineering with Turn-Level Context Injection}

The system employs a dynamic prompt engineering strategy that incorporates both the current system state and summarised outputs from prior turns. This enables the LLM to reason incrementally, preserving context across turns without exceeding token limits. Unlike static, single-shot prompts, this iterative injection approach supports continuous refinement of strategy and improved decision-making based on feedback from the target machine.

\subsection{Enforced Rationale Generation to Improve LLM Reasoning}

To enhance the quality and reliability of each LLM-recommended command, \texttt{PenTest2.0} explicitly instructs the LLM to include a short \textit{rationale} as part of its structured response at every turn. 

By requiring the LLM to explain \textit{why} a command was selected based on the system context and command history, we encourage the LLM to engage in \textbf{reflective reasoning} rather than shallow pattern matching. For instance, in Turn 7 of a representative test run, the LLM explained:
{\scriptsize
\begin{quote}
	\small
	\texttt{"rationale": "The previous command failed due to a syntax error with unmatched quotes. This corrected version uses proper escaping
		for the inner quotes, allowing us to execute `id' non-interactively, while the interactive version spawns a shell for further
		exploration."}
\end{quote}
}
This rationale indicates that the LLM is actively evaluating available binaries, considering historical command outcomes, and adjusting its tactics accordingly.  

In practice, we found that enforcing rationale generation leads to:
\begin{itemize}
	\item \textbf{Better decision-making}: The LLM is less likely to repeat failing commands blindly.
	\item \textbf{Improved transparency}: Human users gain insight into the LLM’s logic and can catch semantic drift or hallucinations early.
	\item \textbf{Easier debugging and auditing}: Rationales provide a lightweight narrative of the LLM's evolving strategy.
\end{itemize}

Thus, rationale generation is an essential mechanism in \texttt{PenTest2.0} to guide, audit, and ultimately improve automated PrivEsc attempts through explainable AI.

\subsection{Root Shell Detection}

\texttt{PenTest2.0} does not rely on the assumption that a successful PrivEsc attempt will result in an explicit \texttt{root} label in the shell prompt. Instead, it applies multiple heuristics and regex-based pattern matching techniques to infer root access — by inspecting outputs of commands such as \texttt{id}, \texttt{whoami}, and shell prompt characteristics. This approach ensures that even in minimal or obfuscated shell environments, the system can accurately determine if root has been achieved.

\subsection{Non-Interactive Command Enforcement for Safe Automation}

During early development, a major challenge emerged: the LLM occasionally suggested interactive commands (e.g., \texttt{sudo passwd} or \texttt{sudo su}) that led to shell hangs or indefinite blocking due to the absence of user input. These situations prevented prompt completion, delayed root detection, and undermined system reliability.

To address this, \texttt{PenTest2.0} introduced explicit prompt instructions directing the LLM to return two variants of each suggested PrivEsc command:

\begin{itemize}
	\item \textbf{Non-Interactive Command:} A command designed for safe, automated execution via SSH. It is syntactically valid, shell-safe, and crafted to perform PrivEsc based on the current system context, without requiring user input.
	\item \textbf{Interactive Variant:} This is a   manually executable alternative to the non-interactive command. This version is typically identical, but may differ in scenarios where interactive input, terminal feedback, or user supervision is desirable. It offers flexibility for manual verification, fallback testing, or deeper analysis.
\end{itemize}

This dual-output strategy significantly enhanced both system robustness and root detection accuracy. By aligning LLM behaviour with automation constraints — while still allowing for HITL    validation when needed, \texttt{PenTest2.0} ensures safe, efficient, and controllable PrivEsc across multi-turn reasoning loops.

\subsection{Human-AI Collaboration}

While the system is designed to operate autonomously, \texttt{PenTest2.0}, as previously stated, includes a mechanism for manual hint injection. This allows human users to steer the LLM by appending natural language suggestions (e.g., ``try sudo misconfigurations'' or ``avoid using the `find' command again''). This collaborative mechanism is particularly helpful in research, debugging, or adversarial benchmarking settings.

In practice, we found that human-injected hints significantly improved the accuracy and efficiency of the system’s responses. They often led to faster root acquisition, reduced repetition of ineffective commands, and steered the LLM away from invalid or resource-intensive suggestions. These results reinforce the value of human-AI collaboration in real-world offensive security applications.

\subsection{Prompt Size and Cost Estimation with User Approval}

In LLMs, the cost of each interaction is directly proportional to the number of tokens in the prompt and response. Tokens are sub-word units used by the model to process input; thus, longer prompts — especially those containing multi-turn reasoning context or injected enhancements — incur higher computational cost. This can quickly translate into significant API charges, particularly in automated systems that loop through multiple reasoning turns.

To mitigate this, \texttt{PenTest2.0} calculates the token count at every turn — immediately before a prompt is submitted to the LLM — and estimates the associated cost based on the selected model’s pricing. This information is displayed to the user in a clear, tabulated format (see, e.g., Fig.~\ref{fig:promptcostnoflags}). The system proceeds with submission only if the user explicitly approves the prompt. This safeguard prevents oversized or bloated prompts that could otherwise result in excessive API charges or premature depletion of available credits.

Indeed, this concern is not merely theoretical. During early internal testing, the absence of this checkpoint led to unintended prompt inflation, rapidly consuming API credits and halting further experimentation. As such, this manual approval mechanism was introduced as a hard-earned lesson. It now serves to enforce cost control, enhance transparency, and ensure that user oversight is preserved throughout — especially important in budget-sensitive or resource-constrained environments.

This approval loop is repeated at every turn, reinforcing HITL  governance before each LLM interaction.

\subsection{Command Execution with User Approval}
We observed  that the behaviour and suggestions of the LLM cannot be fully trusted and may, at times, be unsafe, misleading, or operationally hazardous. For example, the LLM might inadvertently suggest destructive commands such as \texttt{rm -rfv *} (which could wipe out the entire filesystem) or resource-intensive operations like \texttt{zip -rv zipped.zip /} (which attempts to recursively compress the entire system).

To mitigate these risks, \texttt{PenTest2.0}  maintains \textbf{a local blacklist} of dangerous and unsafe commands. If a suggested command matches an entry in this list — either exactly or via pattern matching — it is automatically voided and never executed. Instead, the system flags the command as invalid and proceeds to the next reasoning turn, optionally logging the incident for future review. This safeguard ensures that even if the LLM's reasoning fails or drifts semantically, system integrity is preserved.

Beyond automated filtering, \texttt{PenTest2.0}, as previously stated, requires user approval for each LLM-generated command, shown with its rationale. Approved commands are executed via SSH, and outputs are logged for subsequent reasoning.


\section{Prototype Implementation}
\label{Prototype Implementation}

\subsection{Development Language}
\label{Programming Language}
We developed \texttt{PenTest2.0} in \texttt{Python 3} for its versatility, rich libraries, AI integration, and adaptability, though alternatives like Bash or Go could also have been used.

\subsection{Physical Host and Virtual Environment Configuration}
\label{Physical Host and Virtual Environment Configuration}
The experimental setup was conducted on two physical host machines. Initial experiments were performed on a MacBook Pro equipped with a 2.8 GHz Quad-Core Intel Core i7 processor, 16 GB of RAM, and 1 TB of storage, running VirtualBox 7 for virtualisation. More recent experiments used a Lenovo laptop running Windows 11, powered by an Intel Core Ultra 7 processor, 32 GB of RAM, and 1 TB of storage, also employing VirtualBox 7 to host the virtual environment.

In both cases, the virtual environment comprised the following machines:
\begin{enumerate}
	\item \textbf{Kali Linux VM}: The primary attack platform hosting \texttt{PenTest2.0} for PenTesting.
	\item \textbf{Linux VM}: A 64-bit Debian system with 512 MB RAM, serving as the  main target.
\end{enumerate}
We use a NAT configuration to enable seamless communication between the VMs, effectively simulating a realistic and controlled network environment.

\subsection{GenAI Tool}
\label{GenAI Tool}

\texttt{PenTest++} employs OpenAI's \texttt{ChatGPT} for its advanced capabilities and performance, opting for an online LLM    over a local one to leverage automatic updates and the latest features. Other  GenAI tools, e.g.\  Gemini\footnote{\url{https://gemini.google.com/}} and DeepSeek\footnote{\url{https://chat.deepseek.com/}}, are potential alternatives.

To integrate ChatGPT programmatically with \texttt{PenTest2.0}, we used OpenAI's \texttt{gpt-4o-mini} API, which operates on a subscription-based model\footnote{\url{https://platform.openai.com/}}. This model was selected for its favourable balance between affordability and reasoning performance — offering competitive capabilities as one of the latest LLMs. The current pricing is \$0.15 per million input tokens and \$0.60 per million output tokens\footnote{\url{https://platform.openai.com/docs/pricing}}. While our system supports more advanced models, such as \texttt{o3}, \texttt{gpt-4.1} and \texttt{gpt-4.5}, these were not used due to higher associated costs.

\subsection{Proof-of-Concept implementation}
\label{Proof-of-Concept implementation}
 PenTest2.0  operates as a modular, multi-turn, AI-driven PrivEsc agent. 
We next present  a prototype, proof-of-concept implementation.

\begin{enumerate}
	\item \textbf{Initial Setup:} PenTest2.0 assumes the attacker already has a low-privileged foothold on the target system from a prior exploit. The prototype is executed from a Kali Linux VM, which serves as the attacker’s platform. The user launches the system from the command line by running a Python script, optionally passing flags to enable specific reasoning features (CoT, RAG, PTT, or hint). All runtime parameters are provided in a fully customisable  configuration file that  includes the target machine’s IP address, SSH username and password, the LLM model to be used (e.g., \texttt{gpt-4o-mini}), and the maximum number of reasoning turns (e.g., 10). Upon startup, the system reads and parses these variables, verifies SSH connectivity to the target using the provided credentials, and prepares the execution environment. This setup process includes creating timestamped logging directories, resetting token counters, and loading core modules for command execution, root detection, prompt generation, and other optional enhancements.

 \item \textbf{System Bootstrapping and Reconnaissance:} After the Python script is launched, PenTest2.0 initialises internal modules based on the runtime configuration, enabling optional reasoning features such as RAG, CoT, PTT, and human-authored hint injection if specified. The system then performs remote reconnaissance by executing a predefined set of system probing commands on the target machine via SSH. 
 These include:
 \begin{itemize}
 	\item \texttt{whoami} --- displays the current logged-in username;
 	\item \texttt{id} --- shows the user ID, group ID, and associated group memberships;
 	\item \texttt{hostname} --- reveals the system’s network hostname;
 	\item \texttt{uname -a} --- prints detailed kernel and system architecture information;
 	\item \texttt{cat /etc/os-release} --- provides the OS distribution and version;
 	\item \texttt{uptime} --- reports how long the system has been running along with current load averages;
 	\item \texttt{df -h} --- shows available and used disk space across mounted filesystems in a human-readable format;
 	\item \texttt{free -m} --- displays memory usage statistics in megabytes;
 	\item \texttt{ps aux --sort=-\%mem | head -n 10} --- lists the top 10 processes consuming the most memory;
 	\item \texttt{ss -tulnp} --- lists all open TCP/UDP ports along with the associated processes;
 	\item \texttt{ls -la /home} --- shows detailed contents and permissions of user home directories;
 	\item \texttt{sudo -l} --- lists the commands the current user can execute with sudo privileges;
 	\item \texttt{cat /etc/passwd} --- enumerates system users and their default shells;
 	\item \texttt{cat /etc/group} --- displays defined system groups and their members;
 	\item \texttt{env} --- prints all current environment variables;
 	\item \texttt{ls -la /tmp} and \texttt{ls -la /var/tmp} --- list files in temporary directories often used in PrivEsc; and
 	\item \texttt{find / -perm -4000 -type f 2>/dev/null} --- searches the entire filesystem for SUID binaries that may be exploitable.
 \end{itemize}

  The outputs of these commands are parsed, summarised, and used to construct the initial prompt for the LLM, ensuring it begins reasoning with an accurate and context-aware snapshot of the target environment.

 \item \textbf{Prompt Construction and First Turn:}
 \label{Prompt Construction and First Turn:}
 The system constructs a structured LLM prompt using a predefined multi-part template, embedding the gathered reconnaissance data. The prompt is dynamically assembled using Python logic, where optional components are injected only if explicitly enabled by the user at runtime via command-line flags.

 The full prompt consists of the following  components:
 \begin{itemize}
 	\item \textbf{(i) System Summary:} A concise overview of the target's OS, user identity, privileges, and environment variables, derived from commands like \texttt{id}, \texttt{uname -a}, \texttt{sudo -l}, and \texttt{env}, as detailed above.

 	\item \textbf{(ii) Command History:} This is a  list that tracks attempted commands and their corresponding outputs. While the list may grow during execution, it is capped by a fixed threshold to prevent excessive prompt length. Once this threshold is reached, the oldest entries are discarded to make room for newer ones, ensuring the prompt remains concise and focused on recent context. This design choice guards against prompt bloat and ensures consistent performance. Moreover, PenTest2.0 enforces a maximum number of reasoning turns — defined in the configuration file — which bounds the total number of interactions within each session.

 	\item \textbf{(iii) Reasoning Instructions:} The LLM is instructed to think step-by-step and suggest the next best escalation command. If enabled, Chain-of-Thought (CoT) logic is added to explicitly guide the model through multi-stage reasoning.
 	\item \textbf{(iv) Task-Oriented Goals:} The LLM is instructed to return a compact, syntactically valid JSON object that includes:
 	\begin{itemize}
 		\item a \texttt{non-interactive command} (safe for automated execution) --- this ensures the command can be executed directly via SSH without requiring additional input, enabling hands-free operation and seamless loop integration.
 		
 		\item an \texttt{optional interactive version} (for human verification if needed) --- provides a variant suitable for manual execution in scenarios where the user prefers to test the PrivEsc command interactively, offering a fallback mechanism and aiding future debugging.
 		
 		\item a \texttt{summarised system overview} --- captures essential system facts (e.g., user ID, OS, kernel version) to ground the LLM’s reasoning in the current context and guide informed decision-making.
 		
 		\item a \texttt{short command history} --- maintains a traceable record of recently attempted commands and their results, helping the LLM avoid repetition and enabling coherent evolution across multiple turns.
 		
 		\item a \texttt{concise rationale} (max two sentences) --- justifies the proposed command based on observed system context, offering transparency into the LLM’s reasoning and aiding user understanding.
 	\end{itemize}

 	\item \textbf{(v) Optional Enhancements (Dynamically Controlled):} These components are added to the prompt template \emph{only if explicitly requested by the user} via command-line arguments. Their inclusion is conditionally controlled by Python template logic:
 	\begin{itemize}
 		\item \textbf{CoT:} Adds multi-step guidance for decomposing the reasoning process.
 		\item \textbf{RAG:} Adds context from an external knowledge base (e.g., GTFOBins) using precise search queries generated by the model.
 		\item \textbf{PTT:} Enables task tree tracking with hierarchical IDs, updated statuses, and subtask generation.
 		\item \textbf{Human Hint:} If a human-supplied suggestion is given, it is injected and flagged as a high-priority hint.
 	\end{itemize}
 \end{itemize}
 A critical instruction embedded in the prompt is that the model \emph{must respond with a single, valid JSON object only}. This constraint ensures consistency, safe parsing, and reliable execution within the loop.

 \item \textbf{Prompt Size and Cost Estimation with User Approval:}
 Before the prompt is submitted to the LLM, PenTest2.0 calculates its token count and provides a cost estimate based on the selected model's pricing. This information is presented to the user in a clear, tabulated format (see Fig. \ref{fig:promptcostnoflags}). The system will only proceed with LLM submission if the user explicitly grants approval. This safeguard ensures that the prompt is neither bloated nor unintentionally oversized — conditions that could otherwise lead to excessive API charges or premature budget depletion. 
 Note that this approval process is repeated at every turn, immediately before each prompt is submitted to the LLM.
	
\item \textbf{Command Suggestion and Justification:}
\label{Command Suggestion and Justification:}
	Upon receiving the prompt, the LLM processes the embedded requirements — such as system summary, reasoning goals, and output constraints — and then returns a structured JSON response containing the following key fields:
	
	\begin{itemize}
		\item \textbf{Non-Interactive Command:} A command designed for safe, automated execution via SSH. It is syntactically valid, shell-safe, and crafted to perform PrivEsc based on the current system context. 
		
		\item \textbf{Interactive Variant:} A manually executable alternative to the non-interactive command. This version is typically identical to the non-interactive
		command but may differ in scenarios where interactive input, terminal feedback, or user supervision is desirable. It offers flexibility for manual verification or fallback testing.
		
		\item \textbf{System Summary:} A concise, LLM-generated overview of key system attributes, including the current user, UID, hostname, operating system, kernel version, sudo privileges, and the number of SUID binaries found. This contextual snapshot grounds the model’s reasoning and justifies its command choice.
		
		\item \textbf{Command History:} A summarised list of previously executed commands and their outputs. This helps the model avoid repeating ineffective actions and ensures that each new command is informed by prior context. To prevent prompt inflation, the history is capped to a fixed number of recent entries.
		
		\item \textbf{Rationale:} A brief, context-aware justification — typically one or two sentences — explaining why the proposed command is suitable. For instance, the LLM might highlight that the command leverages an allowed \texttt{NOPASSWD} binary to spawn a root shell via a known shell escape mechanism, making it safe for automated escalation.
		
		\item \textbf{PTT Update (Optional):} If the \textit{PenTest Task Tree (PTT)} feature is enabled, the response includes a \texttt{ptt\_update} field that provides structured task-tracking metadata. This consists of:
		\begin{itemize}
			\item \texttt{initial\_tree:} A hierarchical list of top-level PrivEsc strategies (e.g., \textit{Sudo exploitation}, \textit{SUID binaries}) and their subtasks (e.g., \texttt{P1.3: Exploit sudo-based escalation}). Each task is assigned a unique ID and marked as \texttt{pending}. This block appears only during Turn 1.
			\item \texttt{current\_task\_id:} The ID of the subtask currently being pursued (e.g., \texttt{P1.3}).
			\item \texttt{new\_subtasks:} Any additional subtasks proposed in the current turn (optional).
			\item \texttt{updated\_statuses:} Status updates for existing tasks (e.g., \texttt{done}, \texttt{in\_progress}, or \texttt{skipped}).
			\item \texttt{commands:} A log of executed commands, tagged with task IDs and their outputs.
		\end{itemize}
		This \texttt{ptt\_update} structure is designed to enable  the LLM to persist and reason over evolving goals across multiple turns. To minimise API costs, the entire block is only included when the PTT feature is explicitly enabled using the \texttt{--ptt} flag at runtime.
	\end{itemize}
	
	The LLM response — fully validated as a compact JSON object — provides the foundation for the next step, where the user reviews and optionally approves the proposed command for live execution.

	\item \textbf{User Approval and Live Execution:} PenTest2.0 displays the LLM-generated command and its rationale to the user for manual approval. This dual-approval mechanism ensures safety and ethical oversight. Once approved, the command is executed on the target via SSH, and the output is logged.

\item \textbf{Turn-Based Feedback Loop:} If root access is not achieved (e.g., the output does not contain a \texttt{uid=0(root)} match), the system summarises the command output and proceeds to construct the next prompt. This new prompt incorporates the updated system context, command history, and any relevant task tree adjustments. The process then repeats for a preconfigured number of turns or until PrivEsc is successful.

Both the request and response prompts exchanged with the LLM in subsequent turns follow the same format as used in steps~\ref{Prompt Construction and First Turn:} and~\ref{Command Suggestion and Justification:}, ensuring consistency and traceability throughout operation.

	\item \textbf{Optional Enhancements:} The system supports the following runtime-configurable enhancements:
	\begin{itemize}
		\item \textbf{Local RAG:} Injects relevant knowledge from a local FAISS vector\footnote{In this context, a vector refers to a high-dimensional numerical representation of text — produced by embedding models such as \texttt{text-embedding-3-small} (\url{https://platform.openai.com/docs/guides/embeddings}). Each input sentence is mapped to a dense vector (e.g., of length 1536), where semantic similarity between texts corresponds to geometric closeness in vector space. FAISS enables fast similarity search over such vectors, typically using cosine similarity as the metric.}
		 store populated with markdown content (e.g., GTFOBins).
		\item \textbf{Online GTFOBins:} Pulls real-time online suggestions from the GTFOBins database\footnote{\url{https://gtfobins.github.io/}}.
		\item \textbf{CoT:} Enables intermediate reasoning through few-shot examples (see Section \ref{Background:CoT}).
		\item \textbf{Human Hints:} Allows user-supplied advice to steer LLM behaviour mid-run.
		\item \textbf{PTT Tracking:} The system maintains and updates the PTT with each turn. Each command is tagged with a \texttt{task\_id}, and tasks transition through statuses: \texttt{pending}, \texttt{in\_progress}, \texttt{done}, or \texttt{skipped}. Subtasks may be added dynamically based on the LLM's evolving plan.
	\end{itemize}
	
	\item \textbf{Root Detection and Termination:} After each command execution, the system invokes a shell-root detection routine based on regular expression matching against command output. If root access is confirmed, the loop terminates early; otherwise, it proceeds to the next reasoning turn.
	
	\item \textbf{Logging and Reporting:} Throughout the session, PenTest2.0 logs all prompts, commands, outputs, reasoning steps, token usage, and task tree updates. Upon session completion, it generates structured logs, a session summary, and optional visualisations for audit and research purposes.
\end{enumerate}

\section{Testing Results}
\label{Testing Results}

To assess the robustness and automation capabilities of \textbf{PenTest2.0}, we carried out comprehensive testing across seven distinct full-system configurations. Each setup combined or excluded strategic features such as  Chain-of-Thought reasoning (\texttt{--cot}), human-injected hints (\texttt{--hint}), Retrieval-Augmented Generation (\texttt{--rag}), and the PenTest Task Tree module (\texttt{--ptt}).  All experiments were conducted against a deliberately vulnerable Linux VM pre-configured with PrivEsc vectors.

\subsection{Evaluation Criteria}

We assessed each configuration along the following six dimensions:

\begin{enumerate}
	\item \textbf{Root Achieved}: Did the command ultimately spawn a root shell, regardless of automation?
	\item \textbf{Auto Root Detected}: Did the system autonomously confirm root access via parsed output and terminate accordingly?
	\item \textbf{Turn of Success}: How early in the loop did the system both achieve and recognise root access?
	\item \textbf{Execution Reliability}: Did the LLM adhere to the required structured JSON schema, returning well-formed fields e.g. \texttt{command\_non\_interactive}?
	\item \textbf{Resilience}: Did the system handle malformed inputs, noisy outputs, or interactive shells without crashing or stalling?
	\item \textbf{Self-Healing Capability}: If the LLM initially suggested a syntactically invalid or ineffective command, was it able to adapt in subsequent turns by analysing previous output (e.g., error messages) and generating a corrected, functional alternative?
\end{enumerate}

\subsection{Key Insights}
Table \ref{tab:pentest-summary-final} presents a summary of results across all seven tested configurations. 
\begin{table}[ht]
	\renewcommand{\arraystretch}{1.5}
	\rowcolors{5}{gray!10}{white}
	\setlength{\arrayrulewidth}{1pt}
	
	\begin{tabular}{||c|c|c|c|c|c||}
		\Xhline{2\arrayrulewidth} 
		\rowcolor{gray!30}
		\textbf{\#} & \textbf{Config} & \textbf{Root} & \textbf{Auto-Root} & \textbf{Turns} & \textbf{Notes} \\
		\Xhline{2\arrayrulewidth}
		1 & --cot & \cmark & \cmark & 2 & Auto-root detected; terminated early \\
		2 & --hint & \cmark & \cmark & 2 & ~ \\
		3 & --cot --hint & \cmark & \cmark & 1 & ~ \\
		4 & --cot --hint --rag --ptt & \cmark & \cmark & 2 & ~ \\
		\Xhline{1.5\arrayrulewidth}
		5 & --rag & \cmark & \xmark & 10 & Manual check confirmed root \\
		6 & --ptt & \cmark & \xmark & 10 & ~ \\
		7 & No flags & \cmark & \xmark & 10 & ~ \\
		\Xhline{2\arrayrulewidth} 
	\end{tabular}
	
	\caption{Results of PenTest2.0 across seven configurations (MaxTurn = 10)}
	\label{tab:pentest-summary-final}
\end{table}
 All seven configurations successfully achieved root access, confirmed either automatically or   manually, validating the reliability of PenTest2.0’s command suggestion logic — even under zero-shot, minimal-guidance settings.

However, only four out of seven configurations achieved fully automated root detection, which is critical to PenTest2.0’s design goals. As explained in more
detail below, the discrepancy arises primarily due to execution context and shell behaviour.

\begin{itemize}
	\item In Configurations 1 to 4 in Table \ref{tab:pentest-summary-final}, the LLM consistently returned a syntactically correct \texttt{command\_non\_interactive}, e.g.:
	\begin{verbatim}
		sudo awk `BEGIN {system("id")}'
	\end{verbatim}
	The ShellDetector wrapper then correctly parsed the output (e.g., \texttt{uid=0(root)}) and triggered an early halt, confirming successful escalation.
	
	\item In contrast, Configurations 5 to 7, although issuing correct commands like:
	\begin{verbatim}
		sudo awk `BEGIN {system("/bin/sh")}'
	\end{verbatim}
	failed to detect root automatically. This is because such commands spawn interactive root shells, which the system cannot monitor directly through non-interactive SSH wrappers.
\end{itemize}

Below is an example of an LLM response from a successful system run, demonstrating, in this particular case, strict adherence to the expected format and effective output parsing under optimal conditions.
{\scriptsize
\begin{verbatim}
	{
		"command_non_interactive": "sudo awk `BEGIN {system(\"id\")}'",
		"command_interactive": "sudo awk `BEGIN {system(\"/bin/sh\")}'",
		"system_summary": "- User: naif\n- Sudo: awk\n- Hostname: metasploitable\n...",
		"command_history": "None yet",
		"rationale": "The target has sudo access to awk, which can spawn a root shell.",
		"rag_search_query": "sudo awk PrivEsc GTFOBins",
		"ptt_update": {
			...
			"updated_statuses": [
			{ "task_id": "root_access", "status": "done" }
			],
			"commands": [
			{
				"task_id": "root_access",
				"command": "sudo awk `BEGIN {system(\"id\")}'",
				"result": "uid=0(root) gid=0(root) groups=0(root)"
			}
			]
		}
	}
\end{verbatim}
}

\subsection{Configuration Analysis}
We now present a focused, case-by-case analysis of each system configuration and its performance.
\subsubsection{Configuration 1: \texttt{--cot}}

\begin{figure}
	\centering
	\includegraphics[width=0.7\linewidth]{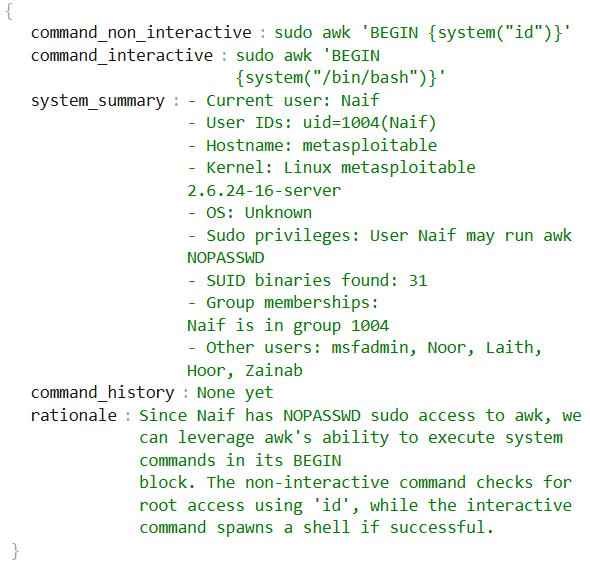}
	\caption{Raw LLM response}
	\label{fig:cotllmresponse}
\end{figure}

This configuration evaluated PenTest2.0 with only CoT reasoning enabled. The system successfully achieved, and auto-confirmed  root access by Turn 2;  the LLM returned a syntactically valid command: \texttt{sudo awk `BEGIN \{system("id")\}'}, which triggered a root shell (see Fig. \ref{fig:cotllmresponse}). PenTest2.0's shell detector correctly parsed the \texttt{uid=0(root)} output and terminated execution early. 

Notably, in other system runs, this same CoT-only configuration achieved root as early as Turn 1. These results underscore a key property of LLM-powered automation: while success is attainable, output may vary across runs, often requiring more than one interaction to converge on the correct result, just as is commonly experienced when manually querying LLMs.

 \subsubsection{Configuration 2: \texttt{--hint}}

\begin{figure}
	\centering
	\includegraphics[width=1\linewidth]{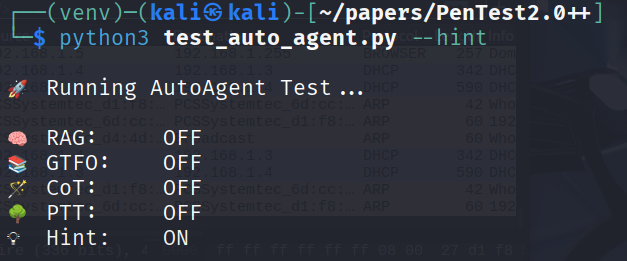}
	\caption{Human hinting feature enabled}
	\label{fig:humanhintenabled}
\end{figure}

\begin{figure}
	\centering
	\includegraphics[width=1\linewidth]{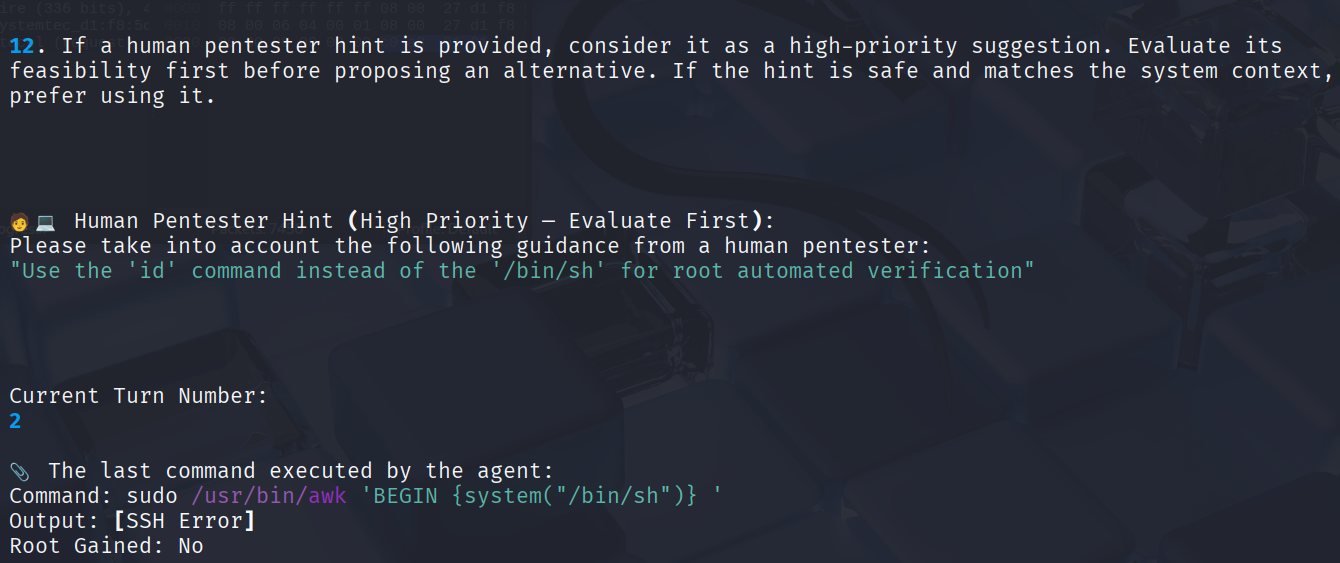}
	\caption{Example of a human-injected hint embedded  in the LLM prompt}
	\label{fig:humanhintembeddedinllmprompt}
\end{figure}

\begin{figure}
	\centering
	\includegraphics[width=1\linewidth]{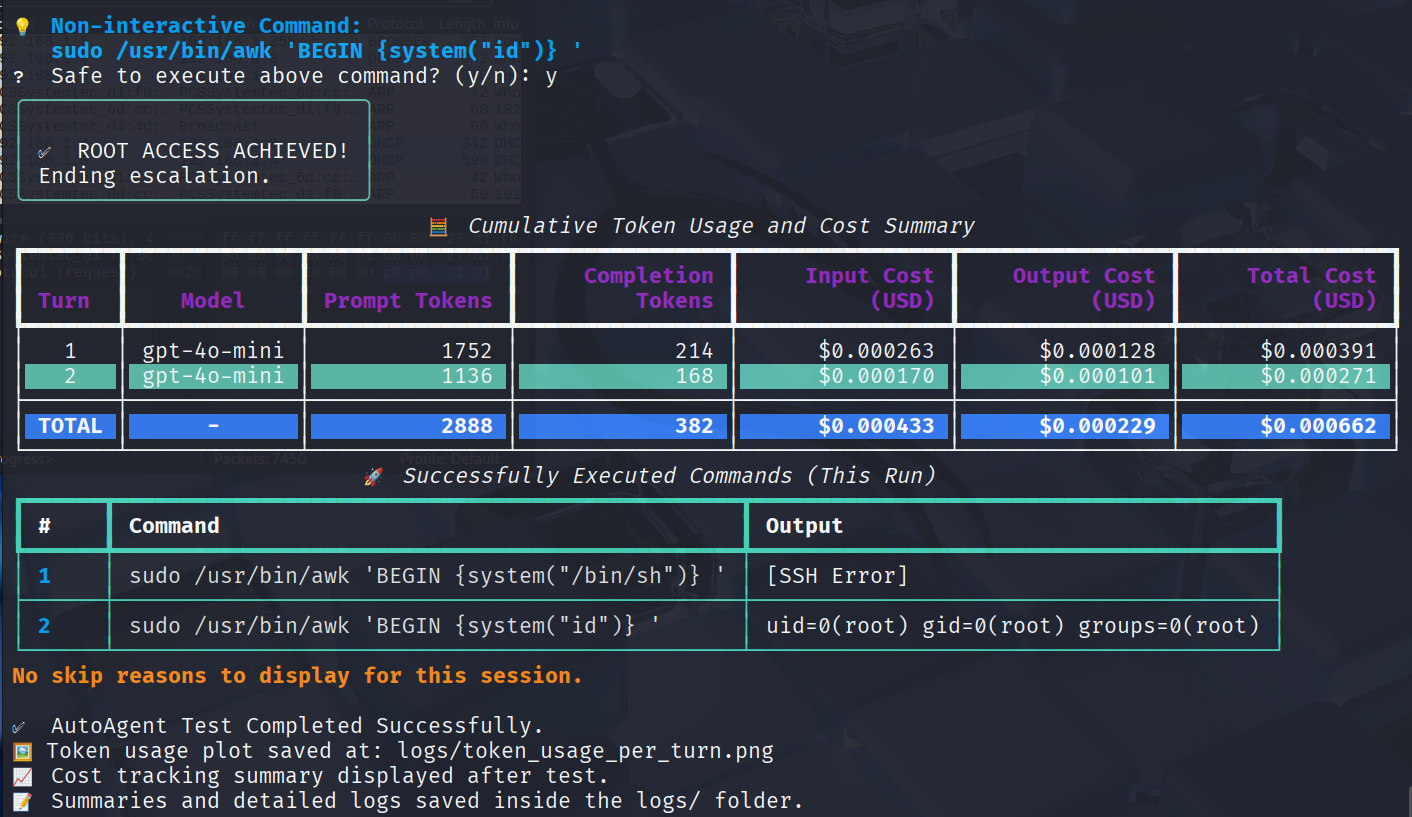}
	\caption{`Root' achieved with the HumanHint feature enabled}
	\label{fig:humanhint-enabledsuccessfulrun}
\end{figure}

\begin{figure}
	\centering
	\includegraphics[width=1\linewidth]{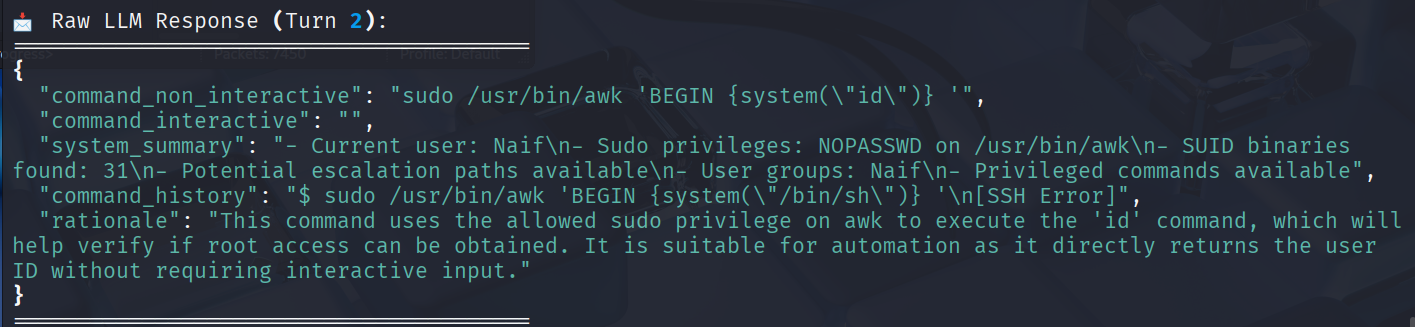}
	\caption{The LLM interprets and responds appropriately to a human-injected hint}
	\label{fig:llmimplementinghumanhintresponse}
\end{figure}

Using only human-injected hints (without CoT, RAG, or PTT), this configuration also achieved root consistently by Turn 2. Automatic root detection worked reliably (see Figs. \ref{fig:humanhintenabled} to 	\ref{fig:llmimplementinghumanhintresponse}). The hints helped steer the LLM without increasing prompt size significantly. Notably, the user was not permitted to inject a hint in Turn 1; this initial turn was reserved to observe whether the LLM could propose a correct command unaided. The hint mechanism was only activated from Turn 2 onwards, allowing the system to prioritise autonomous reasoning before external guidance.

\subsubsection{Configuration 3: \texttt{--cot --hint}}
\begin{figure}
	\centering
	\includegraphics[width=1\linewidth]{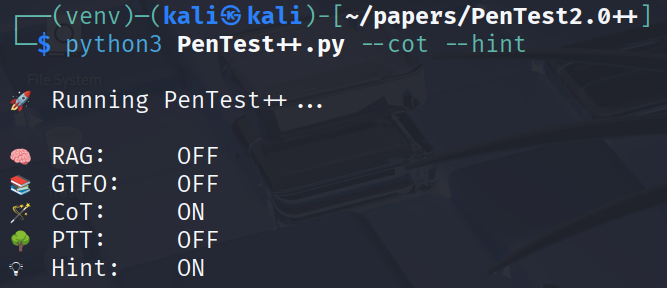}
	\caption{CoT + HumanHint features enabled}
	\label{fig:cothinton}
\end{figure}

This configuration combined CoT reasoning with human-injected hints (see Fig. \ref{fig:cothinton}). The system was executed once and successfully achieved root access in the very first turn. This configuration thus produced the fastest overall result, achieving root in Turn 1, with minimal overhead, no hallucinated commands, and exceptionally low cost.  As designed, the hint mechanism was activated only from Turn 2 onward, allowing the system to first attempt reasoning independently, using CoT, before introducing external guidance.  

We believe that this might be the most powerful, yet cost-effective combination for achieving accurate results. It leverages the  strengths of LLM reasoning, machine training, and human expertise, thereby  advocating for a HITL model. 

\subsubsection{Configuration 4: \texttt{--cot --hint --rag --ptt}}

\begin{figure}
	\centering
	\includegraphics[width=1\linewidth]{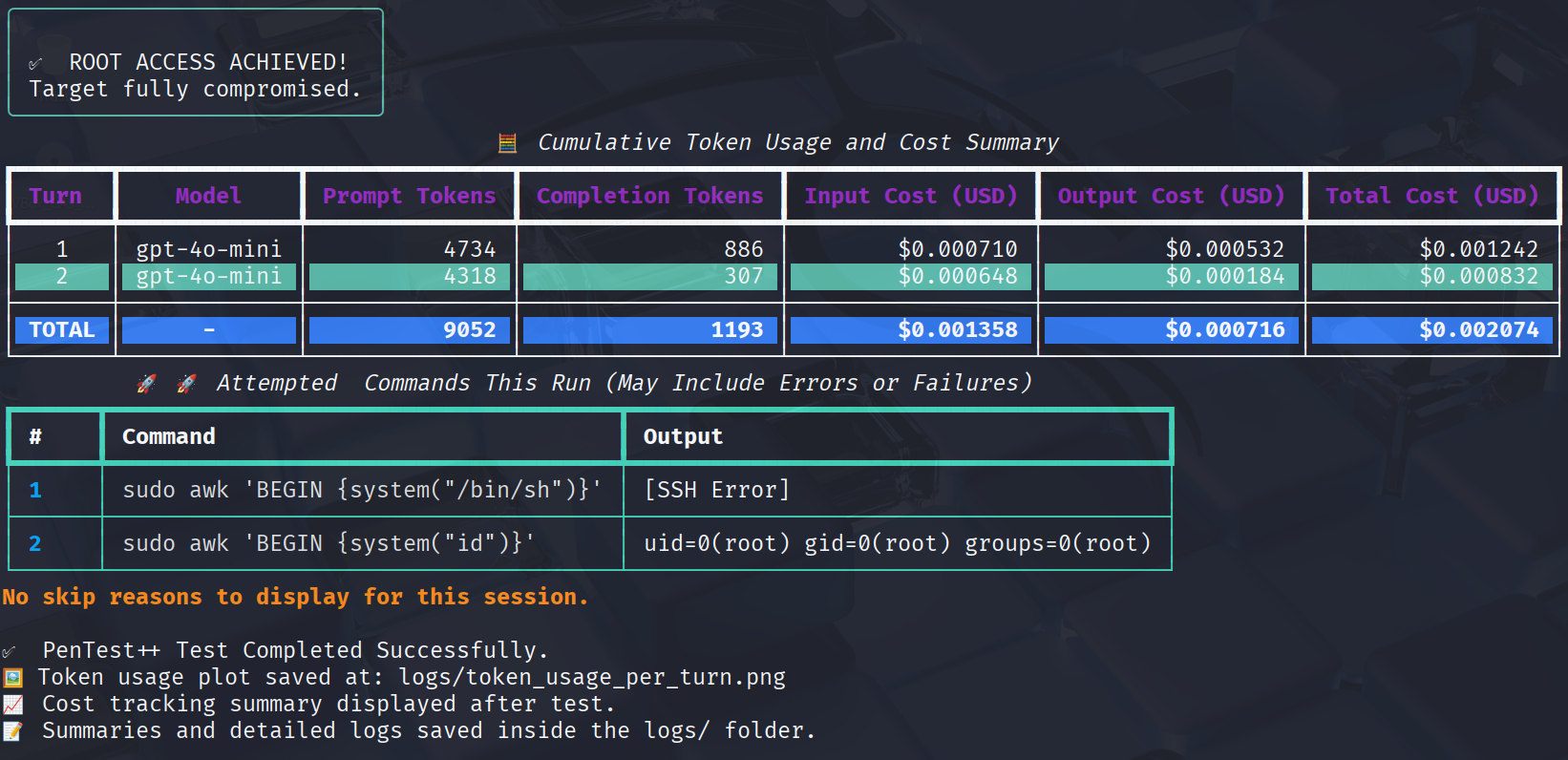}
	\caption{`Root' achieved and auto-detected in Turn 2 using full configuration; cost and outcome tables summarised}
	\label{fig:all-enabledsuccesstablescost}
\end{figure}

\begin{figure}
	\centering
	\includegraphics[width=1\linewidth]{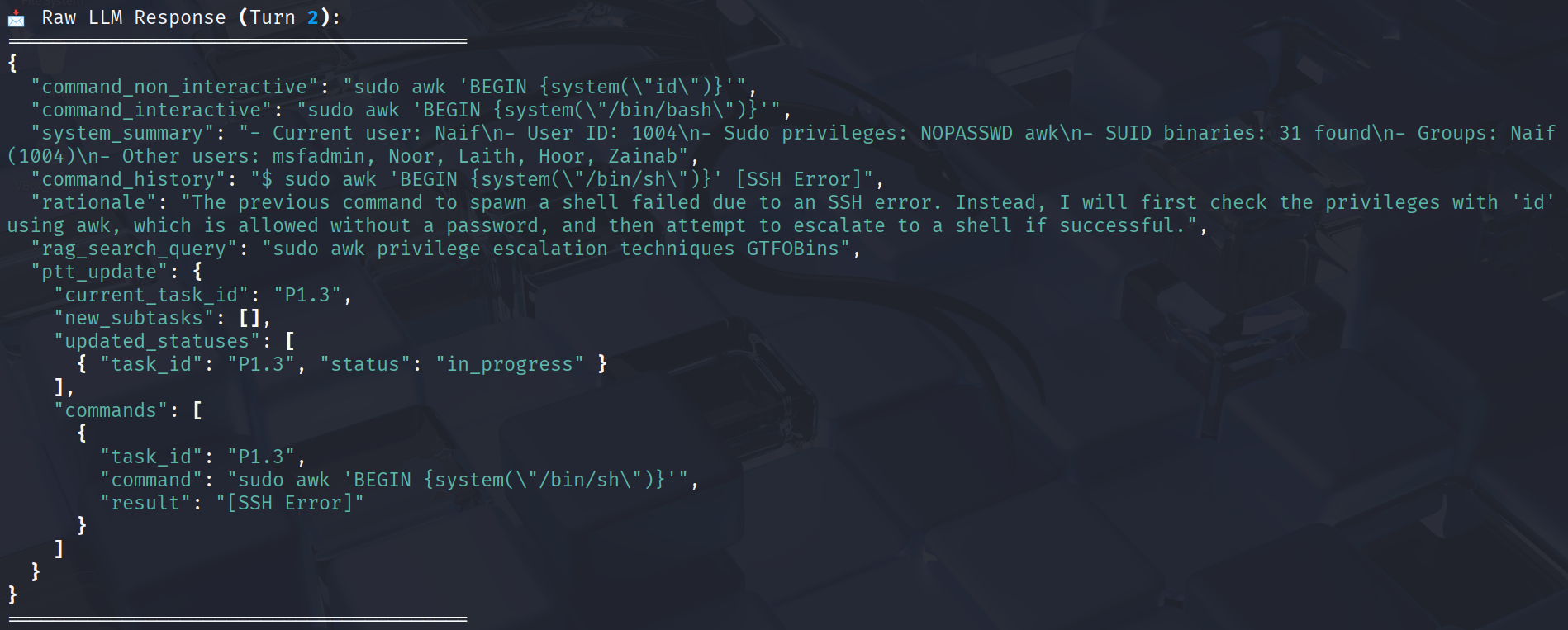}
	\caption{LLM response in Turn 2 showing correct reasoning, format compliance, and command suggestion under the full configuration}
	\label{fig:llm-responseallturn2}
\end{figure}

This configuration enabled all major modules — CoT reasoning, human-injected hints, RAG, and the PTT. Despite the added complexity, the system achieved and detected root access by Turn 2 (see Figs 	\ref{fig:all-enabledsuccesstablescost} and 	\ref{fig:llm-responseallturn2}). The PTT was correctly updated with task statuses and command mappings, and RAG content was successfully injected. However, in this scenario, the added value of RAG and PTT was limited, as the task could be completed with less contextual enrichment.

It goes without saying that this was the most expensive configuration among the seven tested, primarily due to the significant increase in token usage from combining verbose reasoning, injected hints, RAG content, and PTT tracking.


\subsubsection{Configuration 5: \texttt{--rag}}

\begin{figure}
	\centering
	\includegraphics[width=1\linewidth]{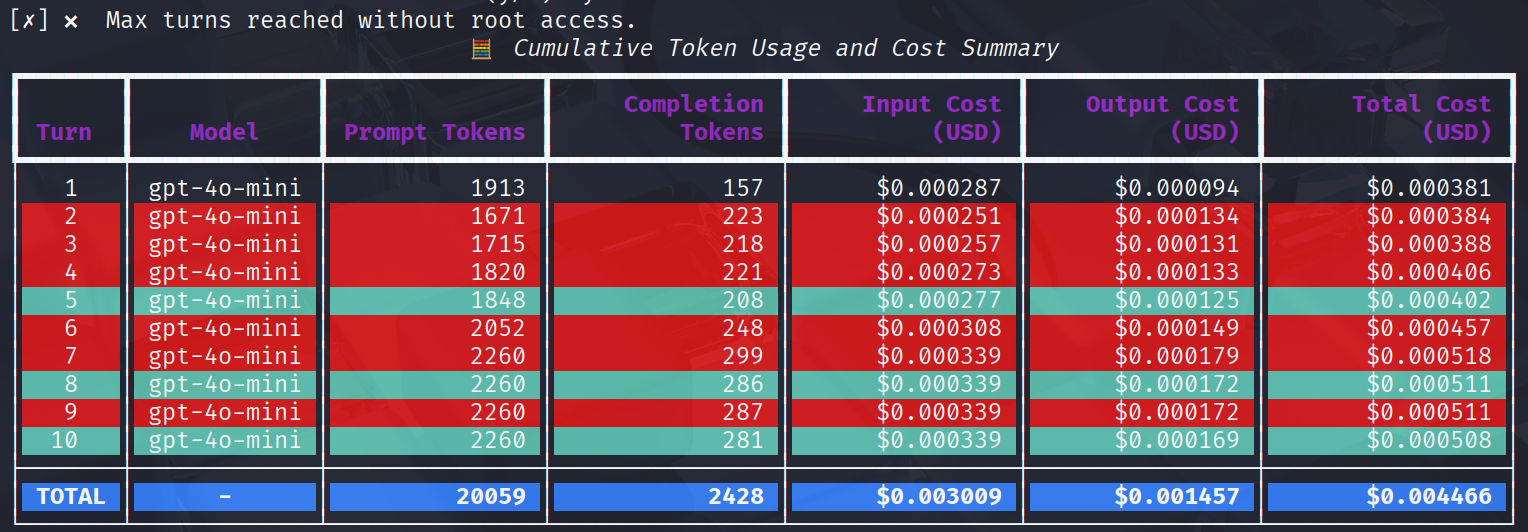}
	\caption{Token cost distribution across failed RAG-enabled runs}
	\label{fig:ragcostsummaryfailure}
\end{figure}

\begin{figure}
	\centering
	\includegraphics[width=1\linewidth]{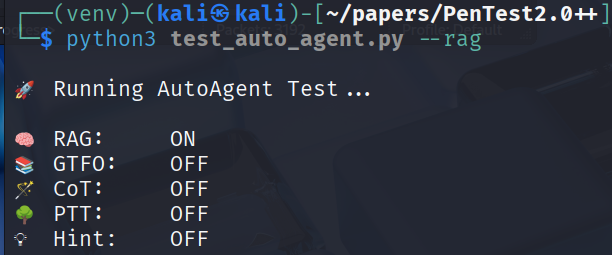}
	\caption{RAG-enabled Configuration}
	\label{fig:ragenabled}
\end{figure}

\begin{figure}
	\centering
	\includegraphics[width=1\linewidth]{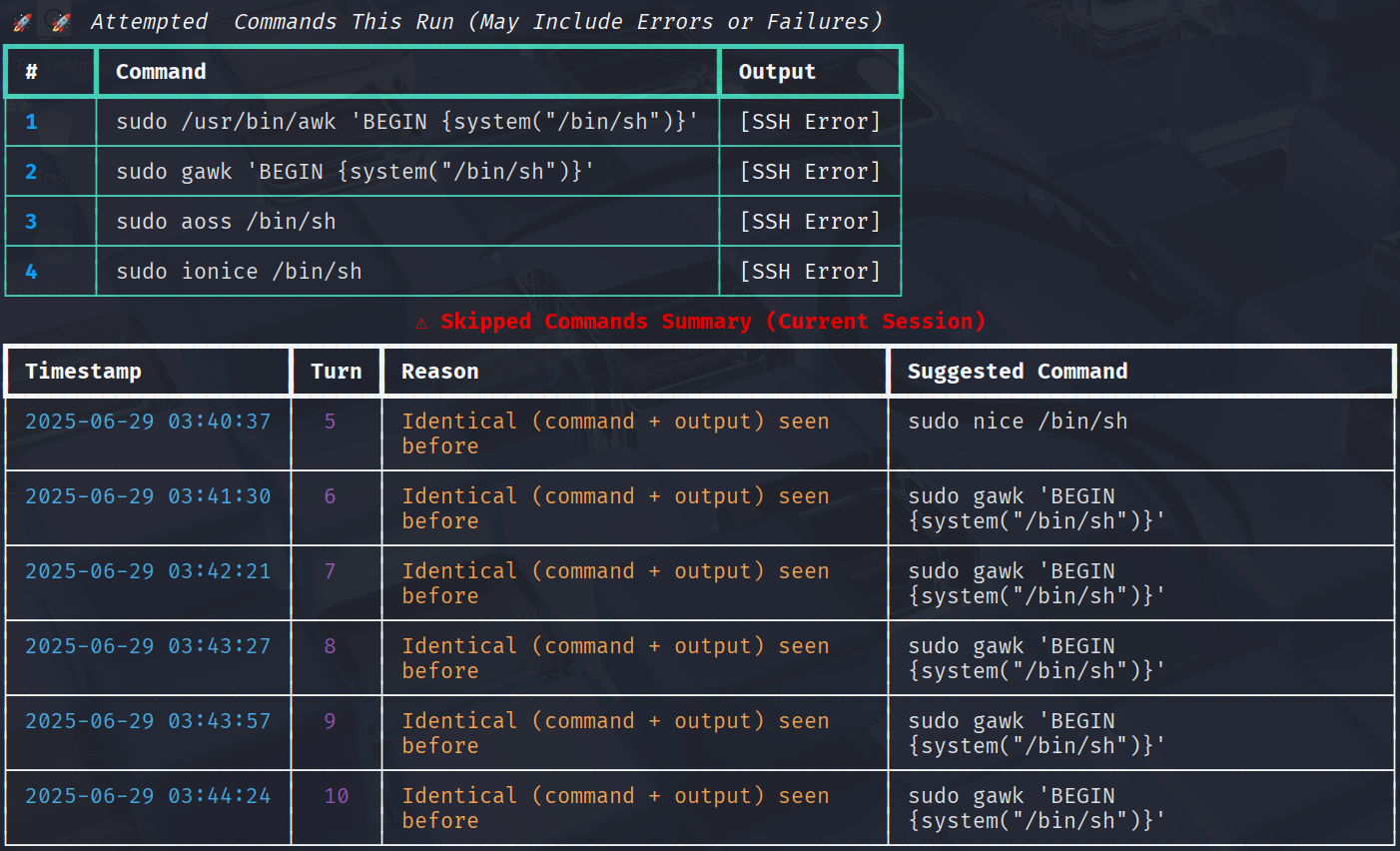}
	\caption{Comparison of repeating vs non-repeating commands during RAG runs}
	\label{fig:ragrepeatingvsnon-repeatingcommands}
\end{figure}

\begin{figure}
	\centering
	\includegraphics[width=1\linewidth]{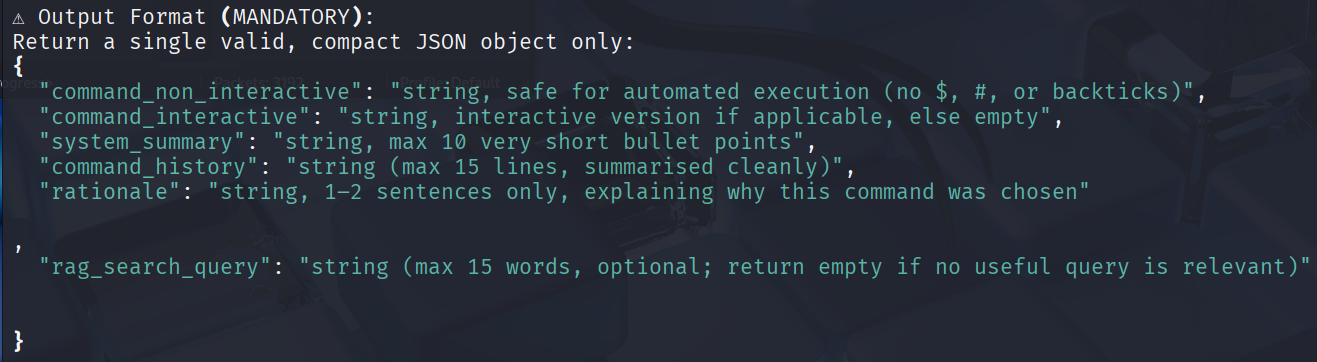}
	\caption{The LLM is instructed to return a RAG  search query}
	\label{fig:ragsearchquery}
\end{figure}

\begin{figure}
	\centering
	\includegraphics[width=1\linewidth]{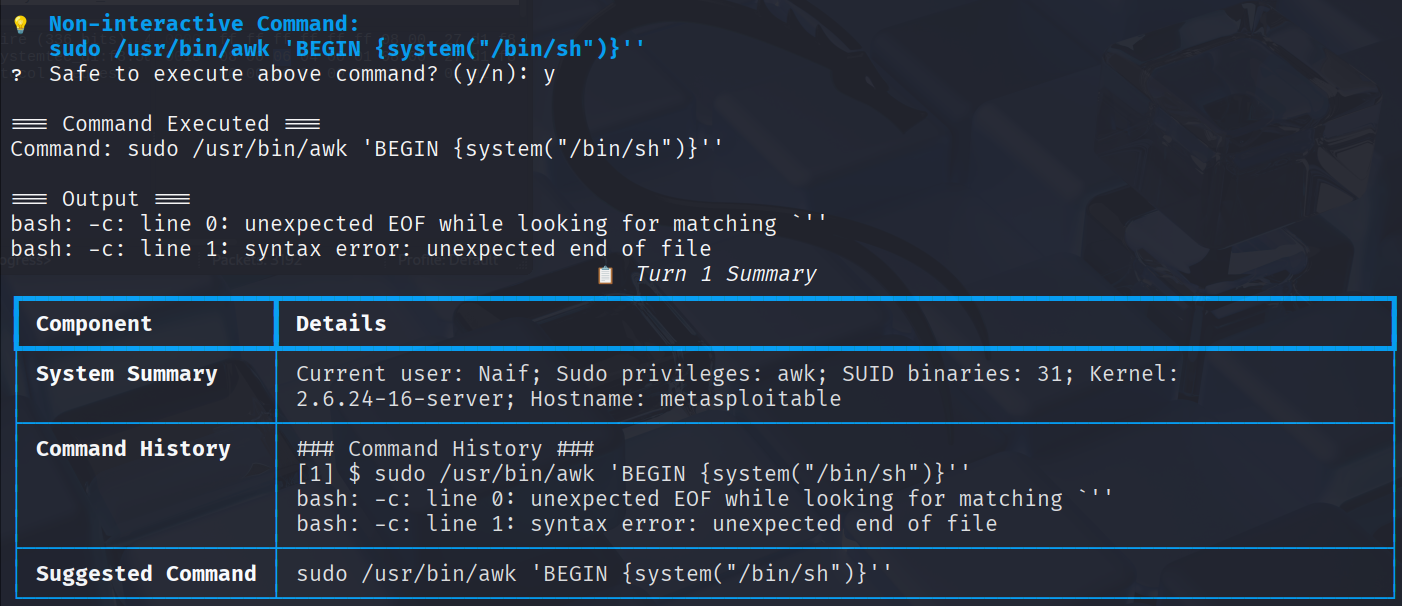}
	\caption{Turn 1 summary showing system context, command history, and suggested command}
	\label{fig:ragturn1summary}
\end{figure}

When using RAG alone — without CoT, Hint, or PTT — the system produced valid commands and successfully achieved root manually as early as Turn 1. However, automatic root detection failed in all 10 runs. This was primarily due to the LLM repeatedly suggesting either interactive root shell–spawning commands or incorrect syntax. Since interactive shells prevent the SSH wrapper from capturing and parsing output, PenTest2.0 could not confirm root autonomously (see Figs.~\ref{fig:ragcostsummaryfailure} to \ref{fig:ragturn1summary}).

These results underscore a critical limitation: despite recent advances, LLMs remain prone to imprecision and cannot yet be fully trusted in high-stakes autonomous security tasks. This reinforces the importance of adopting a HITL  model that combines LLM reasoning with expert oversight.

While the LLM was instructed to suggest a RAG search query, the resulting queries — although relevant — were often too generic to produce impactful retrievals (see Fig. \ref{fig:ragsearchquery}). During testing, we found that using the LLM-generated PrivEsc command itself as the RAG query yielded significantly more relevant and targeted content. However, it's important to note that the RAG corpus lacked knowledge on how to verify root status programmatically. As a result, RAG offered valuable support in discovering exploitation techniques but provided little help in auto-confirming root access, which had to be manually verified.

\subsubsection{Configuration 6: \texttt{--ptt}}

\begin{figure}
	\centering
	\includegraphics[width=1\linewidth]{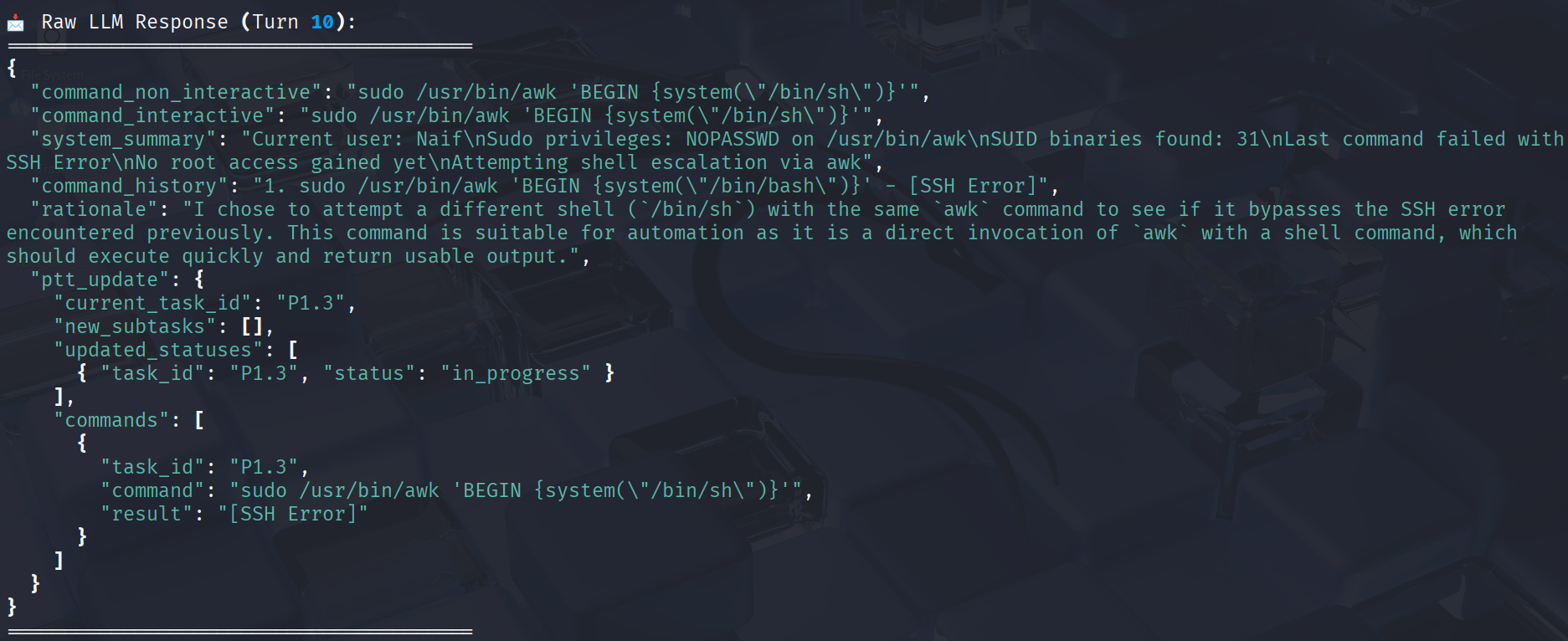}
	\caption{LLM response with structured output in PTT-mode}
	\label{fig:llmresponseptt}
\end{figure}

\begin{figure}
	\centering
	\includegraphics[width=1\linewidth]{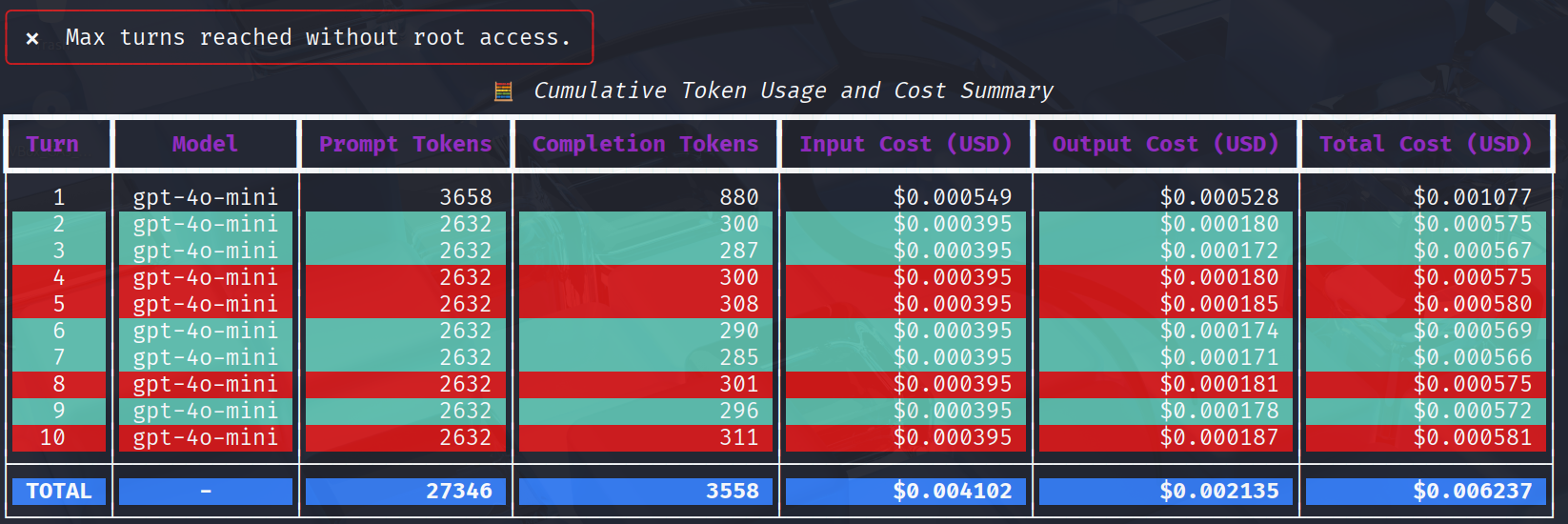}
	\caption{Token cost summary in PTT mode runs}
	\label{fig:pttcostsummaryfailure}
\end{figure}

\begin{figure}
	\centering
	\includegraphics[width=1\linewidth]{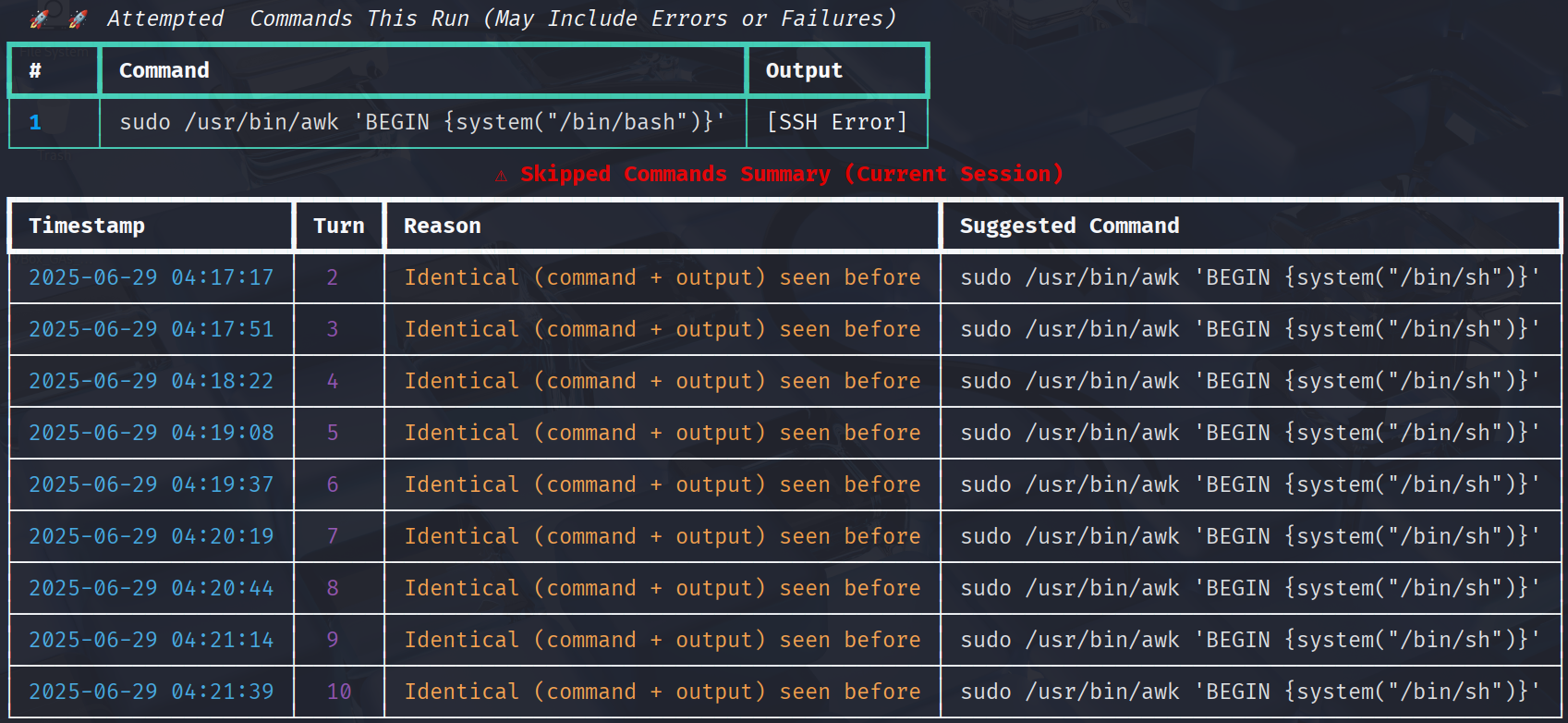}
	\caption{Command repetition vs variation in PTT; high repetition observed}
	\label{fig:pttrepeatingvsnon-repeatingcommands}
\end{figure}

Using only the PTT, the system exhibited good structure and memory of past commands, but it failed to detect root automatically due to the LLM generating interactive shell commands and repeating ineffective strategies (see Figs.~\ref{fig:llmresponseptt} to~\ref{fig:pttrepeatingvsnon-repeatingcommands}). Root was achieved and manually confirmed in Turn~1.

\subsubsection{Configuration 7: No Flags}
With no flags enabled, PenTest2.0 operated in its minimal baseline mode, resulting in the lowest overall token cost (see Fig.~\ref{fig:promptcostnoflags}).
 Although root access was successfully achieved in all ten runs, the system consistently failed to detect root automatically. This was primarily due to the LLM repeatedly suggesting an interactive shell command, despite being explicitly instructed to provide two variants per escalation attempt: a non-interactive version suitable for automation and root detection, and an interactive version for manual testing or debugging.

\begin{figure}
	\centering
	\includegraphics[width=1\linewidth]{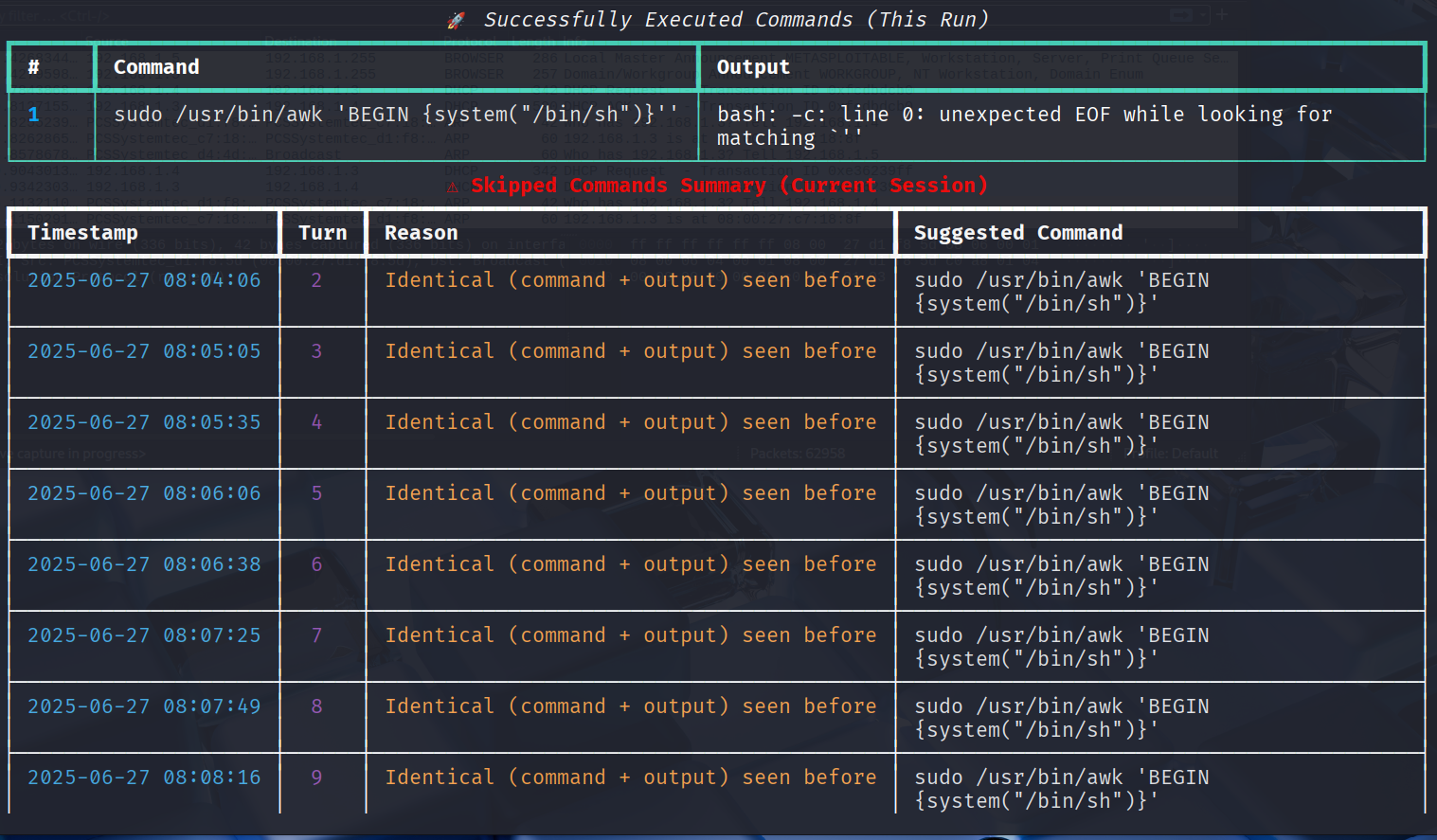}
	\caption{Success vs failure of executed commands}
	\label{fig:successfullyexecutedcommandsvrunsuccessfulones}
\end{figure}

\begin{figure}
	\centering
	\includegraphics[width=1\linewidth]{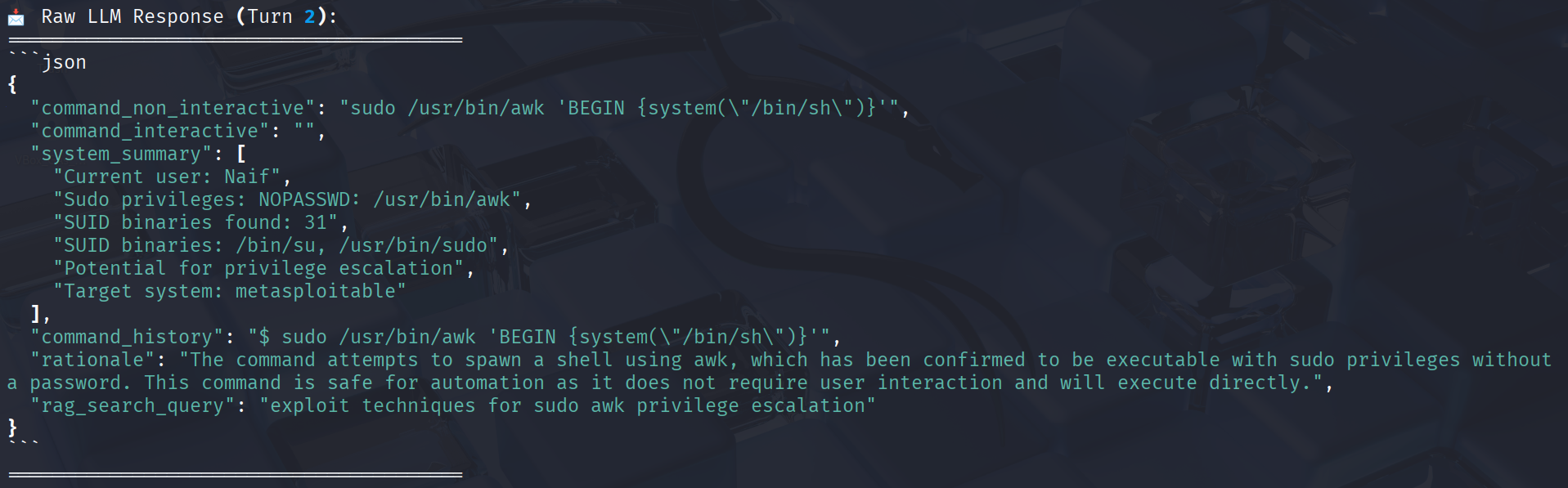}
	\caption{LLM repeating commands despite clear instructions}
	\label{fig:llmnotrespondingwithnewcommanddespiteclearinstructions}
\end{figure}

Furthermore, the prompt clearly instructed the LLM not to repeat previously attempted commands — many of which were already visible in the command history. Nevertheless, the LLM disregarded these constraints and repeatedly proposed the same ineffective strategy across multiple turns (see Figs.~\ref{fig:successfullyexecutedcommandsvrunsuccessfulones}~and~\ref{fig:llmnotrespondingwithnewcommanddespiteclearinstructions}).

This behaviour exposes a fundamental limitation: LLMs may occasionally ignore prompt-level instructions, resulting in non-adaptive and repetitive output. While such lapses are not universal — the LLM does follow instructions correctly in many runs — they reveal a degree of unpredictability that makes full automation unreliable. This reinforces the need for a HITL  architecture in security-critical systems.

\begin{figure}
	\centering
	\includegraphics[width=1\linewidth]{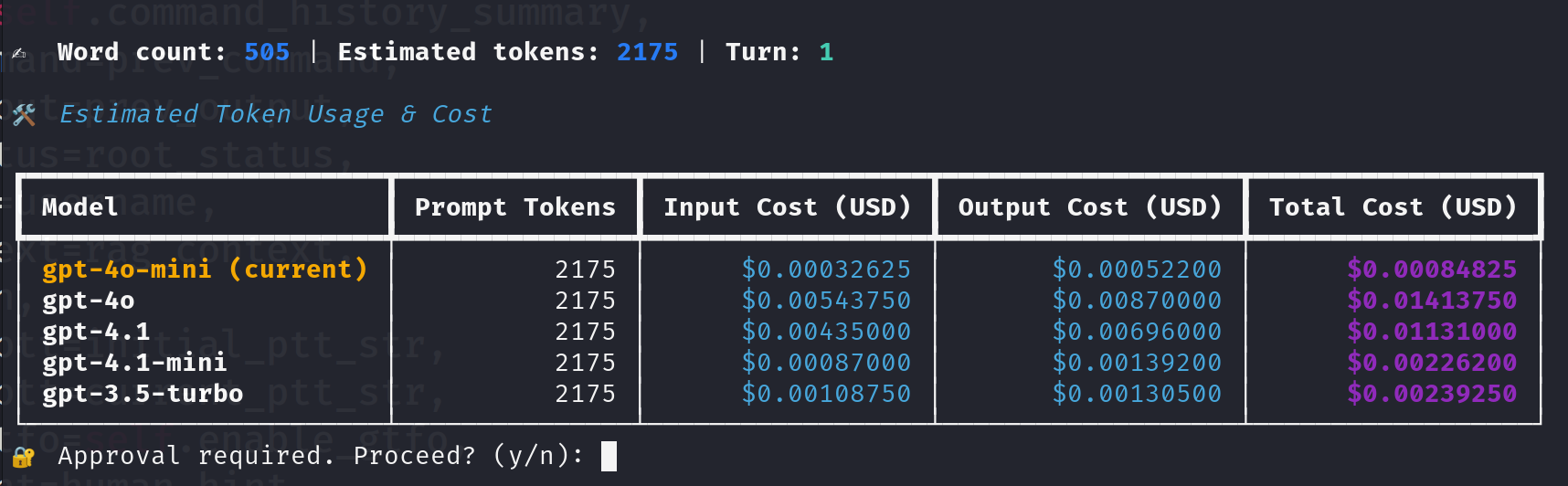}
	\caption{Prompt cost preview shown to user for approval before LLM submission}
	\label{fig:promptcostnoflags}
\end{figure}

\subsection{Comparative Summary}

Table~\ref{tab:config-summary} presents a consolidated overview of the configuration performance across key dimensions, highlighting trade-offs in speed, cost, accuracy, and feature richness.

 \begin{table}[h]
 	\hspace*{-4.5cm} 
 	\renewcommand{\arraystretch}{1.5}
 	\rowcolors{2}{gray!10}{white}
 	\setlength{\tabcolsep}{10pt}
 \begin{tabular}{|c|c|c|}
 	\hline
 	\rowcolor{gray!30}
 	\textbf{Category} & \textbf{Configuration} & \textbf{Remarks} \\
 	\hline
 	Fastest Path to Auto-Root Detection & \texttt{--cot --hint} & Root achieved quickly with minimal  overhead. \\
 	\hline
 	Most Cost-Effective Auto-Root & \texttt{--cot} & Shortest prompts, lowest tokens, fast root success. \\
 	\hline
 	Best Overall Balance (Speed, Accuracy, Cost) & \texttt{--cot --hint} & Robust reasoning and human hints; early root. \\
 	\hline
 	Most Feature-Rich and Knowledge-Intensive & \texttt{--cot --hint --rag --ptt} & All modules enabled: CoT, RAG, hints, PTT. \\
 	\hline
 	Most Expensive Configuration & \texttt{--cot --hint --rag --ptt} & Verbose reasoning and PTT caused high cost. \\
 	\hline
 	Least Expensive Baseline & \texttt{no flags} & Minimalist setup; lowest cost, no enhancements. \\
 	\hline
 \end{tabular}

 	\caption{Summary of configuration performance across speed, cost, and feature dimensions}
 	\label{tab:config-summary}
 \end{table}

\subsection{Performance Against Evaluation Criteria}

To conclude this section, we revisit the six evaluation criteria outlined earlier and summarise how PenTest2.0 performed across each dimension:

\begin{enumerate}
	\item \textbf{Root Achieved}: All seven configurations ultimately succeeded in obtaining root privileges. This demonstrates the LLM's capacity to reason through PrivEsc even under minimal guidance. However, success in some cases (e.g., RAG-only, PTT-only, or no-flags configurations) required manual verification due to shell type limitations.

	\item \textbf{Auto Root Detected}: Only four configurations — those with \texttt{--cot}, \texttt{--hint}, \texttt{--cot --hint}, and \texttt{--cot --hint --rag --ptt} — achieved automated root detection. The remaining configurations issued interactive shell commands that bypassed the SSH wrapper's detection mechanisms. This highlights the need for command predictability and structured output when designing autonomous agents.
	
	\item \textbf{Turn of Success}: The fastest configuration was \texttt{--cot --hint}, which achieved and detected root in Turn 1. Other successful configurations reached root in Turn 2. By contrast, configurations lacking CoT or hinting generally required subsequent turns, with some repetitions and command stagnation along the way.
	
	\item \textbf{Execution Reliability}: Most configurations exhibited high adherence to the structured response format. Failures primarily occurred when the LLM hallucinated commands or ignored prompt instructions, potentially under prompt stress or when operating with insufficient contextual guidance. Configurations featuring CoT reasoning and/or hinting achieved the highest compliance rates.
	
	\item \textbf{Resilience}: PenTest2.0 remained stable across all tests. It did not crash or stall in response to malformed commands, empty outputs, or shell type mismatches. However, interactive shell responses still represent a parsing blind spot, indicating room for enhancement in system robustness.
	
	\item \textbf{Self-Healing Capability}: Configurations with CoT and/or hinting exhibited clear self-correction behaviour, often revising earlier failed strategies by analysing error messages or prompt history. By contrast, the no-flags and RAG-only configurations struggled with command repetition and non-adaptive output, underscoring the limitations of unguided LLM use in security tasks.

\end{enumerate}

Taken together, these findings reinforce the central insight of this paper: while LLMs demonstrate impressive potential in automating complex PenTesting tasks, they require careful prompt design, reasoning scaffolds, and — in some cases — human oversight to ensure reliability, safety, and cost-effectiveness.

\section{Cost Analysis}
\label{sec:cost-analysis}

\subsection{Methodology}

\begin{figure}
	\centering
	\includegraphics[width=1\linewidth]{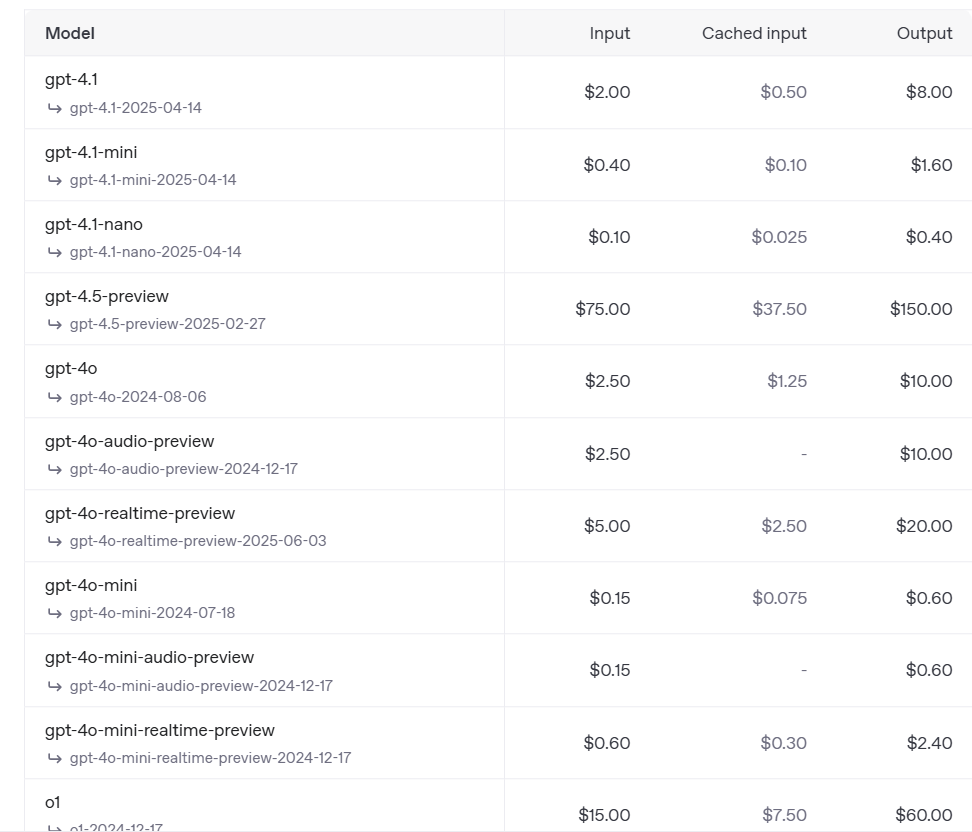}
	\caption{OpenAI pricing for GPT models as of July 2025}
	\label{fig:llmrealcostjuly2025}
\end{figure}

To evaluate the economic viability of each configuration in the PenTest2.0 framework, we implemented a token-accurate cost tracking mechanism based on OpenAI’s pricing model as of July 2025. Each LLM interaction is broken down into prompt (input) tokens and completion (output) tokens. The number of tokens is recorded per turn, and a per-token rate is applied depending on the selected model (e.g., \texttt{gpt-4o-mini}, priced at \$0.15 per 1M prompt tokens and \$0.60 per 1M completion tokens\footnote{\url{https://platform.openai.com/docs/pricing}} --- see Fig. \ref{fig:llmrealcostjuly2025}).

For each session:
\begin{itemize}
	\item \textbf{Input cost} = Prompt tokens $\times$ prompt rate
	\item \textbf{Output cost} = Completion tokens $\times$ completion rate
	\item \textbf{Total cost} = Input cost + Output cost (summed across all turns)
\end{itemize}

Since OpenAI pricing is token-based, PenTest2.0 estimates the number of tokens using a standard approximation: 1 word $\approx$ 1.33 tokens. Accordingly, to compute the number of tokens, PenTest2.0 multiplies the total word count by 1.33. When a prediction is required prior to execution, it also estimates the completion size as 40\% of the prompt size\footnote{\url{https://platform.openai.com/tokenizer}} \footnote{\url{https://github.com/openai/openai-cookbook/blob/main/examples/How_to_count_tokens_with_tiktoken.ipynb}}.

\paragraph{Practical Example.}
Suppose the LLM prompt contains 5{,}000 tokens ($\approx$ 3760 words)  and the expected completion is 2{,}000 tokens (i.e., 40\% of the prompt). Using the \texttt{gpt-4o-mini} model:

\begin{itemize}
	\item Prompt cost = 5{,}000 tokens $\times$ \$0.00000015 = \$0.00075
	\item Completion cost = 2{,}000 tokens $\times$ \$0.00000060 = \$0.00120
	\item Total turn cost = \$0.00075 + \$0.00120 = \textbf{\$0.00195}
\end{itemize}

Over 3 turns with similar token usage, the cumulative cost would be \$0.00585. This level of detail enables PenTest2.0 to estimate, compare, and visualise session costs across configurations with fine-grained accuracy.

\subsection{Configuration-Level Comparison}

Seven configurations were tested under similar conditions. These included simple prompting methods (e.g., \texttt{CoT+Hint}) and complex strategies combining multiple reasoning components (e.g., \texttt{ALL}, \texttt{RAG}, \texttt{PTT}). Each setup was allowed a maximum of 10 turns per session. Fig.~\ref{fig:visual-cost-bar} and~Table \ref{tab:cost-turn-summary}  summarise the cost and turn count distribution.

\begin{table}[h]
	\centering
	\renewcommand{\arraystretch}{1.4}
	\setlength{\tabcolsep}{10pt}
	\begin{tabular}{|c|l|c|c|}
		\hline
		\rowcolor{gray!30}
		\textbf{\#} & \textbf{Configuration} & \textbf{Turns} & \textbf{Total Cost (USD)} \\
		\hline
		1 & CoT + HumanHint & 1  & 0.000533 \\
		2 & HumanHint       & 2  & 0.000662 \\
		3 & CoT             & 2  & 0.000918 \\
		4 & ALL             & 2  & 0.002074 \\
		5 & No Flags        & 10 & 0.003109 \\
		6 & RAG             & 10 & 0.003942 \\
		7 & PTT             & 10 & 0.006237 \\
		\hline
	\end{tabular}
		\caption{Tabular summary of cost and turn count per configuration}
		\label{tab:cost-turn-summary}
\end{table}

\begin{itemize}
	\item \textbf{\texttt{CoT+Hint}} had the lowest total cost (\$0.000533) and completed in a single turn.
	\item \textbf{\texttt{HumanHint}} and \textbf{\texttt{CoT-only}} achieved success in two turns, costing \$0.000662 and \$0.000918, respectively.
	\item \textbf{\texttt{ALL}} incurred moderate cost (\$0.002074) over two turns due to the inclusion of all reasoning layers.
	\item \textbf{\texttt{No-Flags}}, \textbf{\texttt{RAG}}, and \textbf{\texttt{PTT}} reached the 10-turn limit without auto-root detection, costing \$0.003109, \$0.003942, and \$0.006237, respectively.
\end{itemize}

\subsection{Cost-to-Turn Ratio}

\begin{figure}[ht]
	\centering
	\includegraphics[width=1\linewidth]{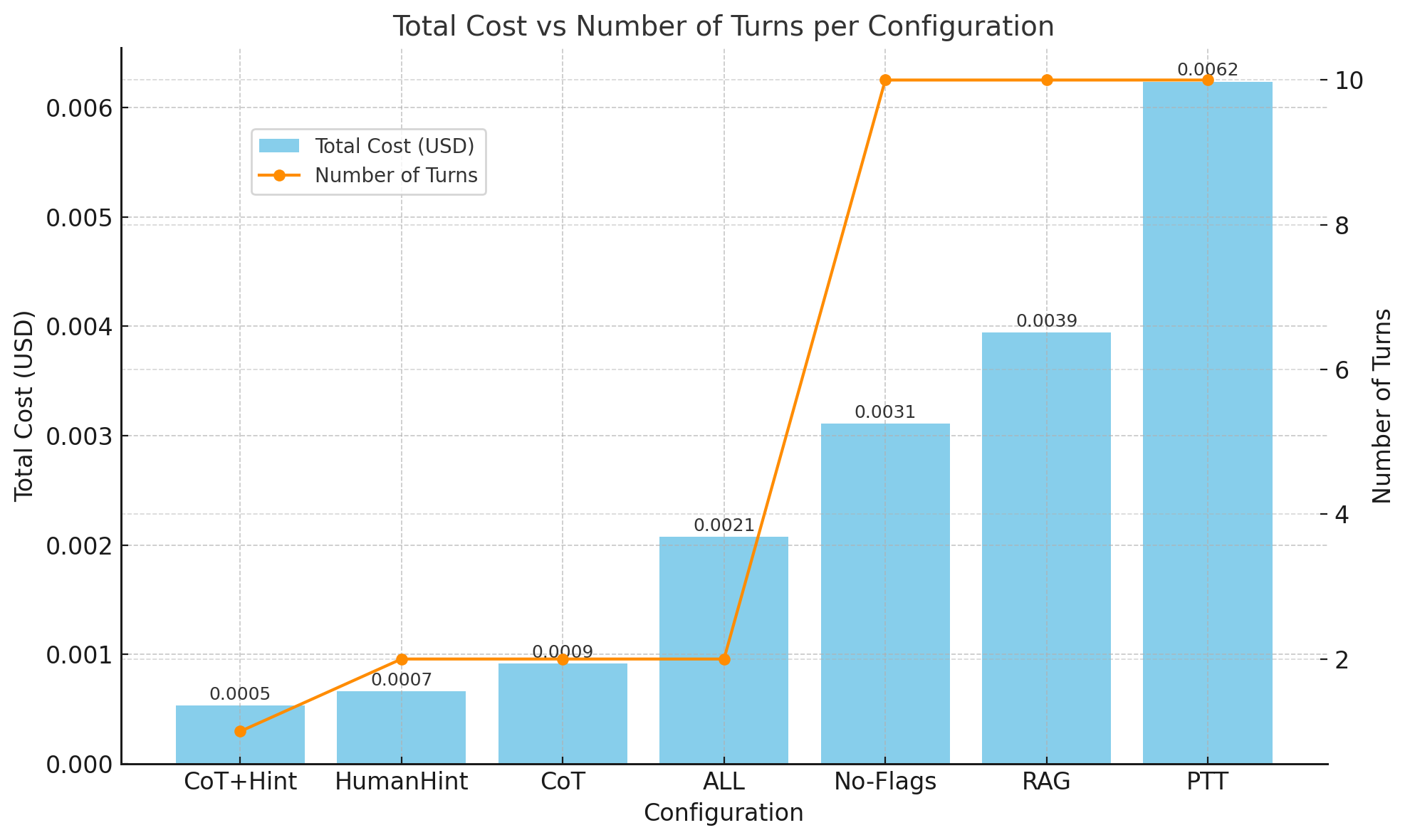}
	\caption{Total cost vs number of turns per configuration}
	\label{fig:visual-cost-bar}
\end{figure}

As depicted in Fig.~\ref{fig:visual-cost-bar}, configurations that succeeded early demonstrated significantly better cost efficiency. The \texttt{CoT+Hint} configuration serves as a strong baseline, achieving success at minimal cost and minimal interaction. Conversely, configurations such as \texttt{PTT}, though theoretically richer in reasoning capability, often generated verbose or repetitive commands, leading to high cumulative cost without better performance.

\subsection{Recommendation and Deployment Guidance}

The evaluation suggests that more expensive configurations do not necessarily yield better results. In fact, verbosity and complexity can lead to degraded efficiency, particularly when interactive shell commands cause failure in automated root detection logic.

We thus advocate a hybrid approach for real-world use:
\begin{itemize}
	\item Use lightweight, low-cost prompt strategies (\texttt{CoT+Hint}, \texttt{HumanHint}) as the default.
	\item Employ more advanced reasoning (e.g., \texttt{PTT}, \texttt{RAG}) only when failures persist or during high-complexity scenarios.
	\item Introduce a HITL  mechanism to intervene when repeated suggestions or ambiguous responses are detected.
\end{itemize}

This pragmatic strategy offers optimal resource usage and ensures a stable cost profile across a range of pentesting scenarios.

\section{General Discussion}
\label{GeneralDiscussion}
\subsection{Answers to the Research Questions}
\label{AnswersToResearchQuestions}

We now address the three research questions (RQs) posed in Section~\ref{Research Questions and Contributions} by integrating  findings from our experimental runs and observations of \texttt{PenTest2.0}’s behaviour.

\begin{enumerate}
	\item \textbf{To what extent can GenAI systems autonomously perform PrivEsc in a post-exploitation scenario under human supervision?}
	Our results demonstrate that GenAI systems — specifically LLMs like \texttt{gpt-4o-mini} — can autonomously identify and execute PrivEsc vectors on realistic Linux targets. In several configurations (e.g., \texttt{--cot}, \texttt{--cot --hint}), the system successfully achieved root access with minimal human input, often within just 1--2 turns. Nevertheless, the requirement for structured prompts and non-interactive shell execution underscores the importance of human supervision in validating command safety and execution context. Thus, while high levels of autonomy are achievable, human oversight remains essential for safety and success verification.
	
	\item \textbf{How can the integration of techniques such as RAG, CoT prompting, PTTs, and optional human hints improve the effectiveness, reasoning depth, and traceability of GenAI-driven PenTesting?}
	Our experiments show that these techniques can enhance LLM performance in meaningful ways. CoT prompting consistently led to faster root discovery and fewer repeated mistakes. Human hinting improved alignment with valid PrivEsc paths, particularly when LLM reasoning stalled or diverged. RAG introduced external contextual knowledge, and PTTs enabled structured tracking of attempted strategies across turns. However, our findings also show that more complex configurations — such as \texttt{--ptt}, \texttt{--rag}, and \texttt{--all} — do not always result in faster auto-root detection. In fact, these setups generally incurred higher token usage and cost due to increased prompt complexity, without guaranteeing proportional improvements in performance. These inefficiencies may stem from several factors, including prompt stress, LLM hallucinations, or architectural limitations — each of which merits deeper investigation in future work. Overall, the \texttt{--cot --hint} configuration offered the most effective balance between efficiency, cost, and success — highlighting the value of guided reasoning and lightweight human-machine collaboration without introducing excessive complexity.

\item \textbf{What are the practical limitations of using LLMs for live, command-executing PrivEsc tasks in realistic environments, and how do these limitations typically manifest during execution?} 
	Several limitations were observed: LLMs occasionally hallucinated unsafe or syntactically incorrect commands, ignored JSON schema formatting, or repeated ineffective strategies despite prior failures. Moreover, interactive shell commands (e.g., spawning \texttt{/bin/sh}) impeded the system’s ability to auto-detect success due to output parsing constraints. Resource-intensive suggestions (e.g., full system compression) also risked destabilising target machines. These behaviours reinforce the necessity of safe prompt design, non-interactive command enforcement, and HITL validation for real-world deployments. A deeper analysis of these LLM shortcomings is provided in Section~\ref{LLMShortcomings}.
	
\end{enumerate}

\subsection{Benefits and Features}
\label{BenefitsPenTest2.0}

\texttt{PenTest2.0} extends the capabilities of its predecessor by introducing autonomous privilege escalation, multi-turn reasoning, and support for advanced LLM-enhanced techniques. The system balances automation, safety, and flexibility — pushing the boundary of what GenAI can achieve in post-exploitation scenarios. Key benefits include the following. 

\begin{itemize}
	\item \textbf{Autonomous PrivEsc with Oversight:}
	\texttt{PenTest2.0} automates the PrivEsc phase of ethical hacking, enabling GenAI to reason, generate, and execute escalation commands iteratively. Importantly, it preserves user oversight through controlled, non-interactive execution and optional hint injection.
	
	\item \textbf{Multi-Turn GenAI Reasoning Loop:}
	The system adopts a loop-based architecture that allows the LLM to process feedback, revise its hypotheses, and try new commands over multiple turns — emulating the trial-and-error nature of real-world PrivEsc.
	
	\item \textbf{Support for Advanced Prompting Techniques:}
	\texttt{PenTest2.0} optionally integrates Chain-of-Thought (CoT) prompting, Retrieval-Augmented Generation (RAG), and PenTest Task Trees (PTTs), allowing for deeper reasoning, traceability, and context retention across execution turns.
	
	\item \textbf{Lightweight Human Collaboration:}
	A streamlined hint mechanism allows human experts to steer the LLM gently when needed — minimising intervention while improving success rates in difficult scenarios.
	
	\item \textbf{Structured Execution and Robust Logging:}
	The system enforces structured prompts and parses outputs in a format-friendly manner, enabling root detection and task tracking. Full command histories and LLM turn responses are logged for auditability and reproducibility.
	
	\item \textbf{Modular, Extensible, and Open-Source Ready:}
	Built in Python with modular components for logging, command execution, LLM communication, and prompt management, the system is cross-platform and designed for future extension. An open-source release is planned to facilitate community engagement and experimentation.

	\item \textbf{Built-in Safety and Cost Controls:}
	Before each prompt is submitted to the LLM, the system automatically estimates and displays the token usage and associated API cost, helping users avoid excessive charges or premature credit exhaustion. Additionally, all LLM-generated commands are screened against a blacklist of dangerous instructions and require explicit user approval before execution — enforcing a strict HITL  policy to mitigate the risk of accidental harm to target systems.
	
	 \item \textbf{User in Control:}
	Throughout its operation, \texttt{PenTest2.0} prioritises ethical alignment by placing the user firmly in control. All command executions are transparent, with the user empowered to monitor outputs, insert corrections, or halt execution when necessary — ensuring safe, responsible use of GenAI in offensive security contexts.
	
\end{itemize}

\subsection{Limitations and Risks}
\label{LimitationsRisksPenTest2.0}

While \texttt{PenTest2.0} represents a substantial advancement over its predecessor, it also introduces new limitations and risks that merit careful consideration. First, although the system supports automated PrivEsc, it does not guarantee success in all scenarios — particularly in the presence of highly customised or hardened environments. The LLM may generate syntactically valid but operationally ineffective or unsafe commands, especially when faced with unfamiliar or ambiguous system states.

Second, the automation of live command execution, even with safeguards such as user approval and command filtering, carries inherent risks. Malicious or erroneous LLM suggestions could result in service disruption, resource exhaustion, or data integrity issues if executed without scrutiny. This risk is heightened in cases where root detection fails, leading to unnecessary or repeated command execution. However, the built-in safety mechanisms and human oversight features detailed earlier — including command blacklisting, cost estimation, and explicit user approvals — serve to significantly mitigate these risks in practice.

Third, the system's reliance on cloud-hosted LLM introduces privacy and data protection concerns. Although prompts are constructed to minimise sensitive content, any external API interaction must comply with applicable data governance policies and ethical guidelines — particularly when operating on production or real-world systems.

Fourth, the system's performance was evaluated under controlled, assumed-breach scenarios on known Linux environments with intentionally exploitable vectors. This limits generalisability to more diverse infrastructures, such as Windows hosts, air-gapped systems, or targets with strict intrusion detection systems (IDS) in place.

Finally, despite using prompt engineering techniques such as rationale enforcement and structured outputs, PenTest2.0 remains susceptible to LLM limitations including hallucinations, semantic drift, and prompt sensitivity. These factors underscore the continued necessity of HITL oversight to ensure correctness, safety, and ethical adherence during PenTesting workflows.

\subsection{Observed LLM Shortcomings}
\label{LLMShortcomings}

Through extensive experimentation with \texttt{PenTest2.0}, we observed several recurring shortcomings in the behaviour of the underlying LLM   —   in our case, \texttt{gpt-4o-mini}. While the model demonstrates strong reasoning capabilities in many instances, its limitations pose practical and operational concerns in an automated PenTesting setting.

First, the LLM can generate erroneous or ineffective commands that fail to escalate privileges, even when the target clearly supports a viable exploit path. In some runs, the model repeatedly suggested syntactically valid commands that produced no meaningful outcome — wasting precious reasoning turns and failing to adapt. This inability to self-correct undermines the system’s autonomy and necessitates human intervention.

Second, we observed that the LLM occasionally suggests dangerously resource-intensive commands, such as recursively compressing the entire filesystem (e.g., \texttt{zip -rv zipped.zip /}). On limited-resource VMs, such commands can degrade performance or crash the system entirely — an unacceptable outcome in real-world testing environments.

Third, the LLM sometimes ignores strict prompt instructions or system guidelines, proposing commands that violate execution constraints or contradict previously observed feedback. This includes bypassing non-interactivity requirements or hallucinating sudo permissions that do not exist. Such deviations underscore the need for robust filtering and manual approval checkpoints.

Fourth, a persistent issue involves the LLM redundantly suggesting the same ineffective command across multiple turns, even after it has already failed (see Figs \ref{fig:ragrepeatingvsnon-repeatingcommands} and \ref{fig:pttrepeatingvsnon-repeatingcommands}). This behaviour leads not only to wasted reasoning opportunities but also to the unnecessary consumption of API tokens — driving up operational costs without progress toward root access.

Collectively, these issues highlight a fundamental limitation of relying solely on LLMs for critical security tasks. They reinforce the importance of a HITL  model, where human expertise governs execution and validates AI-generated outputs. While our observations are based on \texttt{gpt-4o-mini}, more advanced models — such as GPT-4o or GPT-4.5 — may offer improved reasoning, adaptability, and adherence to guidelines. However, these benefits would come at the cost of significantly higher API usage (see Fig.~\ref{fig:llmrealcostjuly2025} ), which may not be feasible in all deployment contexts. Future work should explore this trade-off and assess whether newer models can deliver sufficiently improved reliability to justify the additional expense.

\section{Related Work}
\label{Related work}
 The integration of AI in cybersecurity is an active research area, covering intrusion detection and offensive security, including ethical hacking. Despite significant advances \cite{PSTPaper_PenTestModel_2010,ManualAndAutomatedPenetrationTesting_2016,AutomatedPenetrationTestingAnOverview_2018,2023_GettingGettingPwndbyAI_PenTestingWithLLMs_Happe,hassanin2024comprehensive,2024_PentestGPT_Deng,2025_Penterep_lazarov}, a fully autonomous,  comprehensive  PenTesting system remains elusive.

PentestGPT~\cite{2024_PentestGPT_Deng}  has been proposed as an LLM-powered PenTesting assistant that leverages \texttt{Reasoning, Generation, and Parsing Modules} for a segmented problem-solving strategy. It follows a HITL approach, requiring users to manually enter the target IP, execute suggested commands, and iteratively provide feedback for further guidance. Notably, PentestGPT heavily utilises the concept of PTT, a lightweight memory structure designed to preserve task context, track progress, and reduce reasoning drift across LLM turns. While PentestGPT aims to address context loss, recent content bias, and hallucinations, its reliance on manual execution greatly limits automation. In contrast, \texttt{PenTest++}~\cite{2025_PenTest++_HC_CyBAI} automates predefined PenTesting commands, requiring only user selection and approval. Building on this, \texttt{PenTest2.0} extends automation into the post-exploitation phase, enabling iterative PrivEsc through LLM-guided reasoning, autonomous command execution over SSH, and structured multi-turn feedback loops.
While both PentestGPT and \texttt{PenTest2.0} leverage task-tracking mechanisms such as PTTs, the latter augments them with automatic    cost estimation, command validation, and built-in safety checks — all under explicit user supervision to enforce a robust HITL  model. Ultimately, PentestGPT operates as a capable guided assistant, whereas \texttt{PenTest++} and \texttt{PenTest2.0} pursue deeper autonomy through real-world command execution, proactive risk mitigation, and tighter integration between LLM reasoning and live system interaction.

Our own earlier research explored GenAI’s role in ethical hacking across various phases. In \cite{STM24_UnleashingAIinEthicalHacking}, we proposed a conceptual framework for integrating GenAI into PenTesting workflows. Subsequent studies evaluated ChatGPT’s effectiveness in controlled Windows~\cite{TechReportUnAIInEH_HC_2024} and Linux~\cite{ITASEC25Paper_AdvancingEthicalHackingWthAIALinux-basedExperimentalStudy_2025} environments. More recently, we examined its application in manual exploitation and PrivEsc~\cite{TechReport_APracticalExaminationOfManualExploitationAndPrivilegeEscalationInLinuxEnvironments_HC_2024}. These studies demonstrated GenAI’s potential to enhance efficiency, decision-making, and workflow automation from reconnaissance to reporting.

Building on this foundation, \texttt{PenTest++}~\cite{2025_PenTest++_HC_CyBAI}~\cite{PenTest++_2025pentestelevatingethicalhacking_HaithamChris} introduces a user-centric, AI-powered automation system to streamline PenTesting while maintaining human oversight.
Extending this line of work, the present study introduces \texttt{PenTest2.0}, which advances into the post-exploitation phase by supporting autonomous PrivEsc. It incorporates iterative LLM reasoning, real-time command execution via SSH, automatic output parsing, root detection, and token cost management — all within a HITL framework. This marks a shift from guided task assistance to controlled yet autonomous decision-making and execution within live environments, bridging the gap between LLM-driven reasoning and real-world system interaction.

\section{Conclusions and Further Research}
\label{ConclusionsAndFurtherResearch}

In this paper, we presented \texttt{PenTest2.0}, a significant advancement over our earlier \texttt{PenTest++} system, with a specific focus on automating the PrivEsc phase of ethical hacking. We designed and implemented a GenAI-powered prototype capable of reasoning over multiple turns, generating and executing commands in real time, and autonomously adapting its strategy based on system feedback. Through rigorous experimental evaluation across diverse configurations — including CoT, RAG, PTT, and human hinting — we demonstrated the feasibility of safe, controlled AI-driven PrivEsc in post-exploitation scenarios.

The system leverages structured task tracking, token-aware prompt generation, built-in safety filters, and a human-in-the-loop model to mitigate risks associated with hallucinated or unsafe commands. Our results showed that guided reasoning and hinting (e.g., \texttt{--cot --hint}) yielded the best balance of speed, reliability, and cost. However, our findings also highlighted limitations in LLM behaviour — such as hallucinations, prompt fatigue, and repeated failures — which reinforce the need for user oversight and adaptive error handling.

\texttt{PenTest2.0} marks an important step toward more autonomous and dependable GenAI-assisted PenTesting, particularly in high-stakes, real-world-like environments. By automating PrivEsc while keeping the user in control, it bridges the gap between intelligent assistance and operational autonomy.

Future work will focus on broader evaluation and system expansion. While the current study explores a wide range of configuration options, it is limited to a single target OS environment, which may constrain the generalisability of findings. Testing across diverse systems and vulnerability types will help validate the robustness and adaptability of PenTest2.0 in real-world scenarios. We also plan to extend PenTest2.0 to support additional post-exploitation tasks, including credential dumping, lateral movement, and persistence. Broader environment support (e.g., Active Directory, IoT, macOS, and cloud-native systems) is also a key goal.

Quantitative studies, e.g.   measuring time saved, user trust, and resource efficiency,  will complement our current findings. Additionally, comparative benchmarking with tools like PentestGPT will help establish standardised evaluation baselines. Finally, ongoing work will explore offline LLM deployment, prompt-hardening strategies, and formal risk mitigation frameworks to ensure safe and ethical use of GenAI in penetration testing.

\section*{Acknowledgements}
Some portions of this manuscript were refined using GenAI tools (specifically, ChatGPT) to assist with language polishing and structural clarity. Following the use of such GenAI tools, the authors thoroughly reviewed and edited the content as necessary and take full responsibility for the final publication. All core ideas, findings, experiments, and arguments were developed and authored by the named contributors.

\bibliographystyle{splncs04}
\bibliography{../../../database}

\begin{figure}
	\centering
	\includegraphics[width=0.7\linewidth]{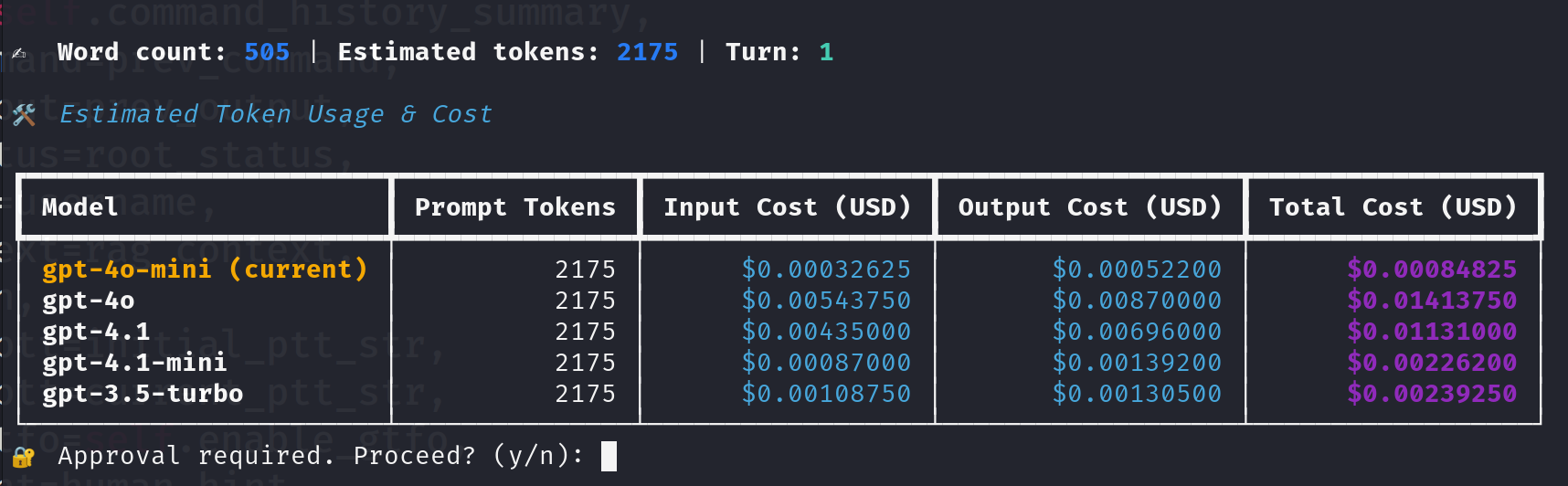}
	\caption{Prompt cost preview with no optional flags enabled}
	\label{fig:promptcostnoflags}
\end{figure}

\end{document}